\documentclass[manuscript,screen]{acmart}
\AtBeginDocument{%
  }

\setcopyright{none}
\copyrightyear{2026}
\acmYear{2026}
\acmDOI{XXXXXXX.XXXXXXX}
\acmISBN{978-1-4503-XXXX-X/2025/06}

\graphicspath{{./figs/}}
\DeclareGraphicsExtensions{.pdf,.jpeg,.png,.eps,.svg}

\usepackage{subcaption}
\usepackage{booktabs, tabularx}
\usepackage{multirow}
\usepackage{xcolor}
\usepackage{xspace}
\usepackage{float}

\usepackage{algorithm}
\usepackage{algpseudocode}
\floatname{algorithm}{Procedure}

\usepackage{listings}
\usepackage{enumitem}
\usepackage{fancybox}  
\usepackage{multirow}
\usepackage{pifont}

\DeclareTextFontCommand{\emp}{\bfseries}

\usepackage[most]{tcolorbox}
\definecolor{custom-gray}{cmyk}{0, 0, 0, 0.7, 1.00}

\newtcolorbox{Summary}[2][]{
top=0.15in,
fonttitle=\bfseries,
colbacktitle=custom-gray,
colback=gray!5,
colframe=gray!40!black,
enhanced,
attach boxed title to top left={xshift=1.5em,yshift=-\tcboxedtitleheight/2},
boxed title style={size=small,colback=custom-gray},
drop shadow={black!50!white},
title=#2,#1}

\begin{document}

\title{Agentic Software Engineering: Foundational Pillars and a Research Roadmap}


\author{Ahmed E. Hassan}
\email{ahmed@cs.queensu.ca}
\orcid{0000-0001-7749-5513}
\affiliation{%
  \institution{Queen's University}
  \city{Kingston}
  \state{ON}
  \country{Canada}
}

\author{Hao Li}
\email{hao.li@queensu.ca}
\orcid{0000-0003-4468-5972}
\affiliation{%
  \institution{Queen's University}
  \city{Kingston}
  \state{ON}
  \country{Canada}
}

\author{Dayi Lin}
\email{dayi.lin@queensu.ca}
\orcid{0000-0002-4034-6650}
\affiliation{%
  \institution{Queen's University}
  \city{Kingston}
  \state{ON}
  \country{Canada}
}

\author{Bram Adams}
\email{bram.adams@queensu.ca}
\orcid{0000-0001-7213-4006}
\affiliation{%
  \institution{Queen's University}
  \city{Kingston}
  \state{ON}
  \country{Canada}
}

\author{Tse-Hsun Chen}
\email{peterc@encs.concordia.ca}
\orcid{0000-0003-4027-0905}
\affiliation{%
  \institution{Concordia University}
  \city{Montreal}
  \country{Canada}}

\author{Yutaro Kashiwa}
\email{yutaro.kashiwa@is.naist.jp}
\orcid{0000-0002-9633-7577}
\affiliation{%
  \institution{Nara Institute of Science and Technology}
  \city{Ikoma}
  \country{Japan}
}

\author{Dong Qiu}
\email{dong.qiu@huawei.com}
\orcid{0000-0001-6255-2618}
\affiliation{%
  \institution{Huawei Canada}
  \city{Kingston}
  \state{ON}
  \country{Canada}
}


\renewcommand{\shortauthors}{Hassan et al.}

\begin{abstract}
Agentic Software Engineering (SE 3.0) represents a new era where intelligent agents are tasked not with simple code generation, but with achieving complex, goal-oriented SE objectives. To harness these new capabilities while ensuring trustworthiness, we must recognize a fundamental duality within the SE field in the Agentic SE era, comprising two symbiotic modalities: SE for Humans and SE for Agents. This duality demands a radical reimagining of the foundational pillars of SE (actors, processes, tools, and artifacts), which manifest differently in each modality. This new vision of SE requires two distinct, purpose-built workbenches (aka tools) for these two collaborative modalities: the Agent Command Environment (ACE), a command center where humans orchestrate, mentor, and oversee agent teams while managing an inbox of agent-generated events like Merge-Readiness Packs (MRPs) and Consultation Request Packs (CRPs); and the Agent Execution Environment (AEE), a digital workbench where agents not only execute tasks but can proactively invoke human expertise when facing complex trade-offs or ambiguity. This bidirectional partnership, which supports agent-initiated human callbacks and handovers, gives rise to new, structured engineering activities (i.e., processes) that redefine human-AI collaboration, elevating the practice from agentic coding to true agentic software engineering. This paper presents the Structured Agentic Software Engineering (SASE) vision, outlining several of the foundational pillars for the future of SE. The paper culminates in a research roadmap that identifies a few key challenges and opportunities while briefly discussing the resulting impact of this future on SE education. Our goal is not to offer a definitive solution, but to provide a conceptual scaffold with structured vocabulary to catalyze a community-wide dialogue, pushing the SE community to think beyond its classic, human-centric tenets toward a disciplined, scalable, and trustworthy agentic future.
\end{abstract}



\maketitle

\section{Introduction}\label{sec:introduction}

Powerful autonomous agents (aka AI teammates~\citep{hassan_se3_2024}) can now write, test, and submit code. They have moved Software Engineering (SE) beyond AI-Augmented development (SE 2.0) and toward Agentic Software Engineering (SE 3.0)~\citep{hassan_se3_2024}. Frontier Large Language Models (LLMs) can generate entire micro-applications from short one-off prompts, suggesting unprecedented productivity. Yet building, and more importantly shipping, complex and trustworthy software for many evolving stakeholders requires structured, iterative, and trustworthy SE practices. The SE field therefore faces a fundamental tension between the velocity of automation and the rigor required to build trustworthy software.

Autonomous coding agents (e.g., Google Jules, OpenAI's Codex, Anthropic's Claude Code, and Cognition's Devin) are already responsible for hundreds of thousands of merged pull requests (PRs)~\citep{li2025aiteammates}. Their hyper-productivity, however, reveals a significant ``speed vs. trust'' gap. Recent examinations of agent-generated code and agent-driven PRs, together with our own hands-on experience with leading autonomous agents, show that many agent efforts still fail to meet the quality bar of being truly ``merge-ready.'' They often contain subtle regressions, superficial fixes, or weak engineering hygiene (e.g.,~\citep{zhang2025which}). Each failed check demands human review, and at agentic scale this verification work can overwhelm developers. At the same time, a new class of practitioners is emerging. These super developers report 100x or even 1,000x productivity by mastering the early practices of this agentic era. Their success is a powerful proof-of-concept, but it also raises a familiar challenge.

The core purpose of the SE field has always been to ensure solutions are trustworthy and delivered economically, and much of the SE field exists because we cannot assume that every team is composed of ``super developers.'' The industry has long acknowledged the phenomenon of ``10x developers,'' a small fraction of developers whose impact far exceeds the median~\citep{mockus2022case}. A significant portion of SE, from structured processes like Agile to sophisticated tools like IDEs, is designed to give non-super developers the scaffolding and opportunity to perform at a 10x level. Agentic SE radically reshapes this landscape, moving the conversation beyond 10x to the realm of 100x and even 1,000x productivity while also redefining the characteristics of such top-tier developers, away from raw coding prowess and toward effective collaboration with fleets of agents (aka AI teammates).

As the industry forges ahead, a Cambrian explosion of ad-hoc practitioner techniques is emerging. However, these grassroots innovations highlight a vacuum of robust, validated approaches. Current methods, relying heavily on informal, conversational prompting, are inadequate for developing trustworthy large-scale, long-lived software. This informality fails to establish robust processes for reproducibility, auditable artifacts for trust, or a durable mechanism for human-agent collaboration. It keeps the paradigm locked in the realm of 1-to-1 ``agentic coding,'' rather than unlocking the potential of N-to-N ``agentic software engineering'' where teams of humans and agents collaborate at scale. Early attempts to impose order, like the Plan-Do-Assess-Review (PDAR) loop, are an important step but not a complete engineering methodology. This new reality calls for more than incremental adjustments. We need to reconsider the pillars on which the SE field is built: the \textbf{Actors}, the \textbf{Processes} they follow, the \textbf{Tools} they use, and the \textbf{Artifacts} they shape.

This call for structure is not unique to software engineering. Parallel debates are unfolding in education, where frameworks for ``human-AI co-thinking'' are being explored to transform learning~\citep{AISwiss2025}. These frameworks emphasize synergistic partnerships, with humans retaining roles of verification and evaluation while treating AI as an intellectual collaborator. Structured Agentic Software Engineering extends this philosophy to engineering, proposing specific structures and disciplines to make such partnerships succeed at scale.

\begin{figure}[t]
    \centering
    \includegraphics[width=0.9\textwidth]{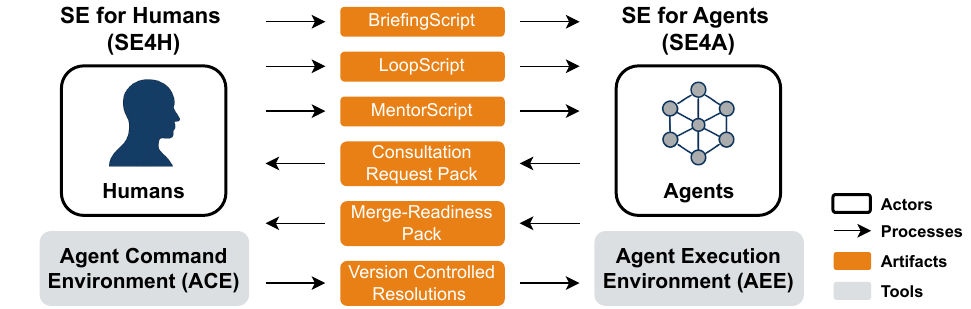}
    \caption{Structured Agentic Software Engineering (SASE) overview.}
    \label{fig:introduction_overview}
    \Description{SASE connects SE for Humans, supported by the Agent Command Environment, with SE for Agents, supported by the Agent Execution Environment. Their shared artifacts are BriefingScript, LoopScript, MentorScript, Consultation Request Packs, Merge-Readiness Packs, and Version-Controlled Resolutions. The framework organizes actors, processes, artifacts, and tools.}
\end{figure}

This paper proposes a vision for this reconsideration: Structured Agentic Software Engineering (SASE), summarized in Fig.~\ref{fig:introduction_overview}. SASE acknowledges that SE is a ``wicked problem'' where rigid, universal processes fail. It therefore prioritizes adaptable solutions, arguing that an AI teammate that can be quickly onboarded into the context of a specific team, project, or organization is more valuable than a brilliant but brittle specialist agent that falters outside its narrow domain. The core thesis of SASE is the introduction of a structured duality, which posits that the field must simultaneously serve two distinct modalities:

\begin{itemize}
    \item \textbf{SE for Humans (SE4H)}, which redefines the human's role to focus on high-level intent, strategy, and mentorship as an Agent Coach.
    \item \textbf{SE for Agents (SE4A)}, which establishes a structured and predictable environment where multiple agents can operate effectively.
\end{itemize}

This duality requires systematically rethinking the four pillars of SE for an agentic era, as they manifest differently in each modality:

\begin{itemize}
    \item \textbf{Actors:} The cast expands from human developers to a hybrid team of human ``Agent Coaches'' and specialized software agents.
    \item \textbf{Processes:} Ad-hoc prompting gives way to structured, repeatable engineering activities that govern human-agent collaborations.
    \item \textbf{Artifacts:} Transient, informal prompts are replaced by durable machine-readable structured artifacts that serve as contracts and institutional memory, including human-authored briefs (BriefingScript) and agent-generated Consultation Request Packs (CRPs) for invoking human expertise.
    \item \textbf{Tools:} The traditional all-in-one human-centric Integrated Development Environment (IDE) is replaced by specialized workbenches designed for the distinct needs and strengths of humans and agents.
\end{itemize}

The traditional IDE is ill-equipped for this new era. As its tooling pillar, SASE proposes two purpose-built environments.

\begin{itemize}
    \item The \textbf{Agent Command Environment (ACE)} is the command center for the human ``Agent Coach.'' It is a workbench optimized for human cognition. It supports strategic tasks such as specifying intent, orchestrating complex workflows, and reviewing evidence-backed results, while offering full observability into agent activities and associated costs.
    \item The \textbf{Agent Execution Environment (AEE)} is the agents' world, a digital workbench optimized for their unique capabilities, such as high-speed computation, massive parallelism, and tireless, repetitive execution, capabilities that far exceed the limitations of human cognition and stamina.
\end{itemize}

Rather than a monologue of informal chat, the interaction between these two environments is a structured dialogue carried by explicit, version-controlled artifacts. Humans initiate the dialogue with the \textbf{BriefingScript} (mission plan), \textbf{LoopScript} (workflow playbook), and \textbf{MentorScript} (best-practices guide), but the exchange is not a static handoff. Agents continue it by generating their own formal artifacts: the \textbf{Consultation Request Pack (CRP)} to request human expertise and the \textbf{Merge-Readiness Pack (MRP)} to present a final, evidence-backed deliverable. Humans then respond with \textbf{Version Controlled Resolutions (VCRs)}, auditable artifacts that formally address each CRP or MRP. Versioned updates to these artifacts capture clarification and feedback over time, keeping the shared understanding of tasks, processes, and team norms current. In this way, SASE turns agentic SE from an informal craft into a disciplined engineering practice.

The presented SASE framework is intentionally visionary. Our primary goal is to provide a conceptual scaffold that catalyzes dialogue throughout the SE community.
SASE is not a temporary scaffold for current AI limitations. As agents become more capable, the central question shifts from what they can do to what organizations can responsibly delegate. Structured human-agent collaboration remains necessary because software goals embody values, priorities, risks, and stakeholder trade-offs that require accountable human ownership. Therefore, trust must rest on evidence, traceability, and auditable decisions. Research on automation~\citep{parasuraman2000model} shows that higher levels of automation often shift risk from execution failure to supervision failure, including loss of situation awareness and fragile handoffs. 

As autonomous agents become first-class actors in the SE lifecycle, the time has come to re-evaluate the foundational tenets of the SE field. We must look beyond the long-held focus on source code as the canonical artifact and the human as the sole actor, and instead build the new processes and tools that are essential for a collaborative, agentic future. This paper is offered as a first step in that collective rethinking, with the express purpose of shaping the discussions that will define the future of the SE field. The paper culminates in a research roadmap that identifies a few key challenges and opportunities, and briefly discusses the implications for SE education.

\section{From Agency to Autonomy: A Hierarchical Framework for AI in SE}\label{sec:background}

To situate the SASE vision within the broader evolution of AI in SE, it is crucial to formalize the progression of intelligent SE (i.e., the integration of AI capabilities into SE). Just as the automotive industry relies on standardized autonomy levels to chart progress in self-driving~\citep{serban2020standard}, we need a comparable framework in SE. We must first distinguish between \textbf{agency}, defined as the capacity of a system to act and execute plans to achieve a given goal, and \textbf{autonomy}, which represents the capacity of a system to self-govern and independently formulate those goals. This distinction allows us to formulate a hierarchical framework, analogous to the Society of Automotive Engineers (SAE) Levels~\citep{sae2021j3016} for autonomous driving, that classifies AI capabilities from simple assistance to full automation. Our presented framework below helps to situate and clarify the transition from AI-Augmented SE (SE 2.0) to the Agentic SE (SE 3.0) era that is the focus of this paper.

\subsection*{Level 0: Manual Coding (No-AI SE) [SE 1.0]}
\begin{itemize}[leftmargin=*,nosep]
    \item \textbf{Canonical Use Case:} No AI mapping. The human manually translates ideas into tokens by typing.
    \item \textbf{Example Technology Manifestations:} Plain text editors like Notepad, vi, and Emacs.
    \item \textbf{Car Autonomy Parallel (SAE Level 0):} No Automation. The human performs all driving tasks.
\end{itemize}

\subsection*{Level 1: Token Assistance (AI-Augmented Coding) [SE 1.5]}

\begin{itemize}[leftmargin=*,nosep]
    \item \textbf{Canonical Use Case:} Maps a developer's immediate editing intent to predicted tokens.
    \item \textbf{Example Technology Manifestations:} Standard auto-complete features in all modern IDEs.
    \item \textbf{Car Autonomy Parallel (SAE Level 1):} Driver Assistance. The vehicle features a single automated system for driver support, such as cruise control.
\end{itemize}

\subsection*{Level 2: Task-Agentic (AI-Augmented SE) [SE 2.0]}

\begin{itemize}[leftmargin=*,nosep]
    \item \textbf{Canonical Use Case:} Maps a planned code change (e.g., a function description) to a complete, generated block of code. Similar levels of automation exist for other SE tasks, like testing and reviewing code.
    \item \textbf{Example Technology Manifestations:} GitHub Copilot, Amazon CodeWhisperer.
    \item \textbf{Car Autonomy Parallel (SAE Level 2):} Partial Automation. The vehicle controls both steering and speed, but the human must constantly supervise and remains responsible.
\end{itemize}

\subsection*{Level 3: Goal-Agentic (Agentic SE) [SE 3.0]}

\begin{itemize}[leftmargin=*,nosep]
    \item \textbf{Canonical Use Case:} Maps a technical goal (e.g., ``add a caching layer'') to a detailed plan of code changes.
    \item \textbf{Example Technology Manifestations:} Emerging agents like Cognition's Devin, Anthropic's Claude Code, Google Jules, and OpenAI's Codex aim for this level. They can take a well-defined goal and execute a multi-step plan (whether self-devised or human-guided) to implement the required changes across code, documentation, and other essential project artifacts.
    \item \textbf{Car Autonomy Parallel (SAE Level 3):} Conditional Automation. The vehicle drives itself under specific conditions, but the driver must be ready to intervene.
\end{itemize}

\subsection*{Level 4: Specialized Domain Autonomy [SE 4.0]}

\begin{itemize}[leftmargin=*,nosep]
    \item \textbf{Canonical Use Case:} Maps a broad technical mandate for a specific domain (e.g., ``ensure the reliability of the payment service'') to a list of concrete technical goals.
    \item \textbf{Example Technology Manifestations:} This level reflects deep, specialized expertise, a frontier that today's most advanced LLMs are beginning to target. Specialization typically occurs along two primary axes: the \textbf{technical stack} and \textbf{quality attributes}. For example, Foundation Models like GPT-5 are now being specialized for the frontend web development domain. As highlighted in GPT-5's official prompting guide, this involves fusing technical skills (such as ``rigorous implementation abilities'' with Next.js and Tailwind CSS) with quality attributes like an ``excellent baseline aesthetic taste.'' By contrast, a Security Agent would specialize along the other axis. It would focus on a single quality attribute (i.e., security) but would be tasked with applying its deep expertise across a diverse range of technology stacks to safeguard all types of software. At Level 4, either axis has to be ensured in one domain.
    \item \textbf{Car Autonomy Parallel (SAE Level 4):} High Driving Automation. A Level 4 car is fully self-driving but restricted to a limited operational domain, such as a ``geo-fenced'' area or, say, specific weather conditions. Similarly, a Level 4 SE agent has high autonomy but only within a particular ``technical domain,'' be it a specific technology stack or a specific quality attribute domain.
\end{itemize}

\subsection*{Level 5: General Domain Autonomy [SE 5.0]}

\begin{itemize}[leftmargin=*,nosep]
    \item \textbf{Canonical Use Case:} Maps a general technical mandate (e.g., ``ensure all our systems are robust'') to domain-specific technical mandates for any unfamiliar domain it encounters.
    \item \textbf{Example Technology Manifestations:} The overarching challenge of domain autonomy lies in scaling this fused capability: consistently achieving deep, specialized expertise across the full spectrum of technology domains (e.g., server backends, embedded systems) and quality attributes (e.g., performance, reliability, accessibility, security), transitioning from the domain-specific Level 4. General domain autonomy is currently at the conceptual/research stage, i.e., it does not yet exist. 
    \item \textbf{Car Autonomy Parallel (SAE Level 5):} Full Driving Automation. A Level 5 car can travel anywhere on any road, in all conditions. Likewise, a Level 5 SE agent can apply its high-autonomy capabilities to any technical challenge, regardless of its technology stack and domain, becoming a truly generalized expert.
\end{itemize}

This framework outlines a trajectory toward \textbf{Autonomous SE} (Levels 4 and 5), but the immediate challenge for industry is the \textbf{Agentic SE} era (aka SE 3.0). The transition from Level 2 to Level 3 is more than an incremental step. It changes the human-computer relationship and introduces new complexity in workflow orchestration, trust, and verification. Before the community can realistically pursue full autonomy, it must establish the disciplined practices needed to manage goal-agentic systems. The SASE vision in this paper therefore focuses on the artifacts, processes, and tools needed to engineer trustworthy software within the SE 3.0 (aka Agentic SE) era.


\section{The Emergence of Agentic Software Engineering}\label{sec:emergence_ase}

\subsection{Industrial Relevance}

The SE field has emerged as a primary proving ground for demonstrating the return on investment (ROI) of large-scale generative AI models like LLMs. This strategic focus by frontier AI labs and the broader industry is motivated by a unique convergence of factors:

\begin{enumerate}[leftmargin=*,nosep]
    \item \textbf{High-Cost Workforce:} Software engineers command premium salaries, meaning even modest productivity gains among this workforce can translate into substantial financial returns. 
    \item \textbf{Rich Training Data:} Code repositories, issue tickets, and commit histories constitute one of the most extensive and well-structured datasets of any knowledge-work domain (a fact the Mining Software Repositories (MSR) community has recognized for decades~\citep{hassan2010msed, hassan2008roadahead}). 
    \item \textbf{Measurable Outcomes:} The field offers clear, quantifiable success metrics (e.g., compiler errors, test outcomes, defect rates) that are critically valuable for creating effective and robust reward functions for reinforcement learning (RL)~\citep{wei2025swe, ma2025sorft}. 
    \item \textbf{Robust Safety Nets:} The presence of automated testing and CI pipelines mitigates the risk that such models will fail in practice. 
    \item \textbf{Transferable Benefits:} Foundation Models honed on SE workflows~\citep{li2025fmse} generalize well to other business tasks that lack equivalent guardrails and data. This powerful combination of factors has, in turn, ignited a cutthroat competition among several frontier companies, all racing to define and dominate the nascent market for Agentic SE.
\end{enumerate}

This concerted industrial focus has rapidly transitioned \textbf{Agentic Software Engineering}, corresponding to the \textbf{Goal-Agentic (Level 3)} stage of our framework, from a theoretical concept to a boardroom-level strategic imperative. This is evidenced by the release of specialized coding agents by all leading players (e.g., \textbf{Google Jules}, \textbf{OpenAI's Codex}, and \textbf{Anthropic's Claude Code}) and a flurry of strategic mergers and acquisitions aimed at securing the invaluable stream of developer feedback data generated from AI-native development environments. This competitive landscape underscores the industry's intense focus on securing the SE data flywheel, which comprises both a user base and the associated feedback data needed to refine the next generation of SE-focused LLMs~\citep{hui2024qwen2coder, wang2024openhands}.

\subsection{What Is an Agent? Key Recent Observations}

Despite the high stakes of Agentic Software Engineering, the fundamental concept of an ``agent'' remains loosely defined. To bring clarity, we situate different agent implementations on a spectrum defined by \textbf{agency} (executing a given plan) and \textbf{autonomy} (formulating the plan itself). This mapping helps clarify the distinction made by frameworks like Anthropic's~\citep{anthropic2024effectiveagents}:

\begin{itemize}[leftmargin=*,nosep]
  \item \textbf{Workflow Agents (High Agency):} These are predefined orchestrations of LLMs and tool invocations. Because their high-level logic and flow are hardcoded by a human developer, they primarily exhibit \textbf{agency} by executing a given plan to achieve a goal. While low-level tasks within the plan might be handled with some autonomy, the overall system is not self-governing.
  \item \textbf{Autonomous Agents (High Autonomy):} These are systems where agents are given a high-level goal and exhibit \textbf{autonomy} by planning, reasoning, and invoking tools to formulate their own path to completion.
\end{itemize}

This distinction is critical. Workflow agents, as systems of agency, require ongoing manual updates to their core orchestration logic to adapt. In contrast, autonomous agents can be iteratively guided and improved by human developers through natural language. This latter approach eliminates low-level code rewrites and enables flexible adaptation, representing a form of ``FMware'' that is coded and rewired using English prose~\citep{cogo2024tutorial}. This marks a fundamental shift in how we build software: from explicitly coding logic to declaratively describing behavior.

\subsection{Brief Survey of Today's Agentic Solutions on Benchmarks like SWE-Bench}

One lesson from the Agentic SE space is that passing tests alone is no longer enough. Recent deeper examinations of SWE-Bench results~\citep{jimenez2024swebench} (the de facto benchmark for evaluating Foundation Model capabilities in SE) highlight a key limitation. The code generated by today's Foundation Models is still far from being merge-ready for professional codebases:

\begin{itemize}[leftmargin=*,nosep]
  \item 29.6\% of ``plausible'' fixes introduced behavioral regressions or were incorrect upon rigorous retesting~\citep{wang2025solved}.
  \item True solve rates for GPT-4 patches dropped from 12.47\% to 3.97\% after detailed manual audits, revealing widespread weak or cosmetic solutions~\citep{aleithan2024swebenchplus}.
  \item AI agents frequently produced superficial patches limited to single files, unlike human developers~\citep{badertdinov2025swerebench}.
  \item Many patches passing unit tests failed broader CI checks due to style or hidden regressions~\citep{xu2025swingarena}.
\end{itemize}

SWE-Bench Verified~\citep{swebench_verified} was introduced to address concerns that some SWE-Bench tasks were ambiguous or underspecified. OpenAI, working with the authors of SWE-Bench, released SWE-Bench Verified as a human-validated subset of 500 tasks. Results on this subset showed rapid progress: GPT-4o solved 33\% of tasks at the August 2024 launch, while leading agentic solutions exceeded 70\% by mid-2025. Yet the benchmark has also become harder to interpret. OpenAI later warned that SWE-Bench Verified was increasingly exposed to data contamination and recommended SWE-Bench Pro~\citep{openai_swebench_pro, swebench_pro} as a more reliable evaluation. The progression from SWE-Bench to SWE-Bench Verified and then to SWE-Bench Pro therefore illustrates both the field's rapid progress and the difficulty of maintaining uncontaminated measures of agent capability.

These benchmark shifts emphasize that passing tests, even on verified tasks, is not enough. Even in mature, production-grade ecosystems like the .NET runtime, developers are observing the same pattern: \textbf{passing tests is far from sufficient}.\footnote{\url{https://www.reddit.com/r/ExperiencedDevs/comments/1krttqo/my_new_hobby_watching_ai_slowly_drive_microsoft/}} Achieving merge-ready status requires a deeper understanding of context, intent, and the broader system. These are qualities that today's agents still struggle to demonstrate reliably.

\subsection{Brief Survey of Today's Agentic Solutions in the Wild Using GitHub Data}

Projections from industry leaders, including Google's Chief Scientist Jeff Dean,\footnote{\url{https://www.businessinsider.com/google-boss-ai-junior-coder-within-a-year-2025-5}} suggest that AI agents will soon perform at the level of junior developers. One early study~\citep{watanabe2025agenticcoding}, which focuses on Claude Code, found that agent-assisted contributions were commonly used for refactoring, documentation, and testing activities. Notably, 83.8\% of these pull requests were eventually merged, and more than half of the accepted contributions were integrated without additional modification.

Recent large-scale studies of GitHub repositories indicate that agentic coding is already widespread. For example, the AIDev dataset~\citep{li2026aidev} contains 932,791 agent-authored pull requests~(PRs) generated by tools such as OpenAI Codex, Devin, GitHub Copilot, Cursor, and Claude Code across 116,211 repositories (dataset cutoff: August 1, 2025), providing one of the first large-scale views of agent participation in software development. Based on AIDev, a large-scale study of 15,451 refactoring instances across 12,256 agent-authored PRs found that agents frequently perform localized and consistency-oriented refactorings, such as variable renaming and type updates, while undertaking fewer high-level architectural changes than human developers~\citep{horikawa2025agenticrefactoring}.

At the same time, several studies highlight important limitations. An analysis of 2,303 agent context files (``Agent READMEs'') from 1,925 repositories found that developers primarily provide agents with functional guidance, such as implementation details, architectural information, and build instructions, while specifying security and performance requirements far less frequently~\citep{chatlatanagulchai2025agentreadmes}. Similarly, a study of 4,550 agent-authored pull requests reported that agents often handle logging and observability inconsistently, with human developers subsequently correcting many of these issues during review and integration~\citep{ouatiti2026logging}.

\section{Motivational Example: The Anatomy of an Agentic SE Workflow}\label{sec:motivation}

\begin{figure}[t]
    \centering
    \includegraphics[width=\linewidth]{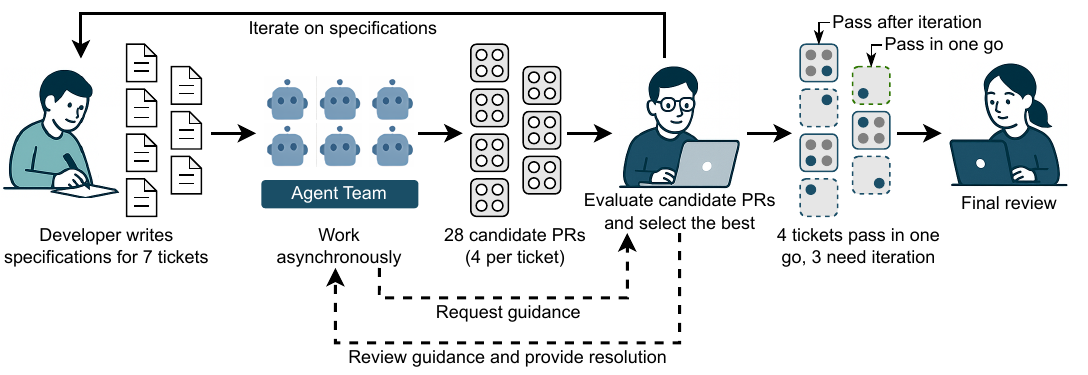}
    \caption{Overview of an agentic SE workflow.}
    \label{fig:example}
    \Description{A developer writes specifications for seven tickets. An agent team works asynchronously to generate 28 candidate pull requests, four per ticket. The developer evaluates the candidates and selects the best. Four tickets pass in one attempt, and three require iteration on the specifications before final review. Agents can request guidance, and the developer provides resolutions.}
\end{figure}

This section grounds the Agentic SE era in a concrete example: a developer resolving seven distinct tickets in a production codebase. This workflow~(as shown in Fig.~\ref{fig:example}) shows both the new capabilities of this era and the process and tooling gaps that SASE is designed to close.

\subsection{The New Workflow: A Glimpse into Agentic SE in Practice}

In this scenario, the developer's role shifts from coder to specifier. Instead of writing code for each ticket, they spend approximately 1.5 hours authoring detailed, natural-language specifications and guidance for each of the seven tickets. These specifications then trigger a team of autonomous agents to work asynchronously, generating 28 distinct pull requests in parallel (4 pull requests per ticket). This highlights the re-emergence of N-version programming~\citep{brilliant1989nversion, chen1995nversion, liew2017nversion, Ron2025Galapagos}, a powerful practice that not only serves as a form of inference-time compute to increase the probability of a successful outcome through trial and error but also enables creative exploration.

For each ticket, the developer evaluates the four solutions and selects the most promising one that satisfies the initial natural-language specification. If none is acceptable, the developer refines the specification and re-triggers the agents. Once all tickets have yielded an acceptable solution, the selected solutions are submitted for review and eventually approved and merged into the codebase.

\subsection{The Process and Artifact Gaps}

This new workflow exposes critical gaps in current SE processes and the artifacts that are used to support them:

\subsubsection{The Art of the Briefing: From Vague Tickets to Actionable BriefingScripts}

A common failure pattern when using AI agents is to simply paste a raw ticket and expect magic. The informal, natural-language tickets are a source of ambiguity. A more rigorous process would treat this specification as a first-class artifact, moving toward structured specifications where the degree of formality can be adapted to the task at hand. The 100x/1,000x developers treat an agent as a junior team member or an outsourced partner rather than a magical tool. They excel at providing a comprehensive initial briefing that includes not only the specification but also the bigger picture, relevant context, and strategic advice on how to break down the task and how to go about doing it.

To formalize this crucial skill, we advocate for moving beyond ad-hoc prompts toward a structured artifact we call a \textbf{BriefingScript}. A BriefingScript is more than a specification of intent. It is the detailed work order that a senior developer would give a junior one, complete with:

\begin{itemize}[leftmargin=*,nosep]
  \item \textbf{What \& Success Criteria:} Defines the scope with a verifiable checklist, similar to Scrum's ``Definition of Done,'' but enriched with formal, testable properties like pre-conditions and invariants.
  \item \textbf{Architectural Context:} Clarifies where the work fits in the system, identifying key modules, data models, or APIs to interact with.
  \item \textbf{Strategic Advice:} Recommends specific implementation approaches, such as libraries to use or patterns to avoid, guiding the agent's problem-solving strategy.
  \item \textbf{Potential `Gotchas':} Highlights known pitfalls or tricky areas to watch out for, like subtle business logic, performance constraints, or dependency issues.
\end{itemize}

Crucially, a BriefingScript is not a rigid, one-shot specification. Like pair programming or collaborative design, it evolves through iterative dialogue between the human coach and the agent. Early drafts may be lightweight, progressively enriched with clarifications and refinements based on agent feedback. This iterative style avoids the brittleness of upfront, waterfall-style specifications and reflects how elite engineers already operate in practice. 
This approach aligns with broader work on human-AI co-thinking~\citep{AISwiss2025}, where humans act as verifiers and evaluators while AI provides generative power. By codifying guidance into BriefingScripts, we ensure that agents operate with both clarity and accountability.
In addition, the BriefingScript must be a living document that evolves over time. Subsequent feedback and clarifications between humans and agents must be incorporated back into the BriefingScript as versioned updates. This approach directly mirrors the principles of managing institutional knowledge in large-scale software engineering. As detailed by Hyrum Wright and others in Software Engineering at Google~\citep{winters2020software}, elite engineering organizations rely on shared, version-controlled knowledge bases as a durable source of truth for human developers. By applying this same proven principle to human-agent collaboration, the BriefingScript transforms from an initial set of instructions into an evolving, auditable record that always reflects the complete and current shared understanding of the task. 

We can make this process more robust by creating version-controlled BriefingScripts. Rather than a new programming language, a BriefingScript is a structured, machine-readable artifact that may be serialized in formats such as Markdown, YAML, JSON, or a domain-specific schema. This approach can be viewed as a modern, agent-oriented evolution of Donald Knuth's concept of ``literate programming.'' Instead of treating code as the primary artifact, the focus shifts to the human-readable BriefingScript: a document that explains the logic and intent from which the agent's work is derived. Looking ahead, creating high-quality BriefingScripts is a significant skill of the elite software engineer, and AI-powered tools could greatly assist developers in drafting them. As an agent becomes more attuned to a codebase and its human collaborator(s), the explicit, human-drafted portion of these scripts should shrink, making the entire SE process more efficient. A concrete example of a BriefingScript is provided in Appendix~\ref{sec:appendix_briefing}.

\subsubsection{The Multidimensional Nature of Agentic Feedback and Mentorship}\label{subsubsec:mentorship}
In the agentic SE era, the feedback and mentorship loop becomes significantly more complex than traditional code review. Guidance is not limited to the final code but extends to the entire SE process, encompassing both explicit instructions and implicit principles that the agent must infer. Key dimensions of this new feedback model include:

\begin{itemize}[leftmargin=*,nosep]
  \item \textbf{Explicit \& Durable Mentorship:} Direct, generalizable guidance from a human coach (e.g., ``Avoid obvious comments; instead, comment on the design rationale'') must be captured durably. This prevents the agent from repeating mistakes and is the motivation for \textbf{MentorScript}, a version-controlled rulebook that codifies best practices.
  \item \textbf{Inferred Mentorship:} Not all guidance is articulated as a general rule. An agent must be able to infer a broader principle from a specific contextual correction by a human, enabling it to learn instead of being spoon-fed. For example, after a human refactors a piece of code for better readability, the agent should propose a new, general rule about that pattern for the coach to approve and add to MentorScript.
  \item \textbf{Holistic Process Feedback:} Mentorship extends beyond the code to encompass the entire SE lifecycle. A human coach may provide feedback on the agent's problem-solving approach, its test planning strategy, the way it debugs or fixes build issues, or its choice and use of tools.
  \item \textbf{Feedback on Multiple Solutions:} The ability of agents to generate multiple potential solutions (\textit{N-versions}) introduces novel feedback patterns. A coach's guidance might involve synthesizing a final solution from different drafts, such as instructing an agent to ``combine the UI from solution 1 with the backend logic from solution 2.''
\end{itemize}

\subsubsection{From Ambiguous Control to Explicit Orchestration}
As observed by industry leaders like Andrej Karpathy,\footnote{\url{https://x.com/karpathy/status/1954224651443544436?s=46&t=ayaQZ2-uUhbu4rg402jW5w}} agents cannot infer the expectations or ``stakes'' of a task. They may ``overthink'' a simple request or under-deliver on a critical one. 
For some tickets, a coach may want to grant the agent full autonomy, while for others, they might wish to enforce a strict process. This process can be defined at an organizational level by process engineers (e.g., for regulatory purposes) and optionally overridden by the coach for a specific ticket, with the override being recorded. Today, this is handled via ad-hoc prompt hacking in the master prompt for these agents.

This motivates the need for \textbf{LoopScript}, a declarative language for defining the agent's workflow, allowing a coach to explicitly communicate the required level of rigor and enforce a precise Standard Operating Procedure (SOP) when needed. We are already seeing this happen in the frontier Autonomous Coding Agents: some agents, like Google Jules, have included a planning step from their inception. Others, like Claude Code, have only recently added an on-demand planning mode in which the agent generates a plan and awaits human review before proceeding.

\subsubsection{From Code Review to Evidence-Based Oversight}\label{subsubsec:merge_pack}
The ultimate goal is to produce a merge-ready contribution. Rather than reviewing dozens of raw pull requests, a human reviewer should focus on auditing a structured \textbf{Merge-Readiness Pack}. This is a bundle of evidence designed to bridge the critical gaps between current agent outputs and the standards of a truly merge-ready contribution. The pack supports the reviewer's assessment by providing evidence for five key criteria:

\textbf{(1) Functional Completeness.} \textit{The Gap:} Agents often produce superficial or partial fixes that pass a narrow set of tests but fail to address the holistic user need. \textit{The Evidence:} The pack provides evidence (e.g., test results) of how the feature behaves in realistic scenarios, allowing the reviewer to assess whether it is complete and meets its specification.

\textbf{(2) Sound Verification.} \textit{The Gap:} Agents may generate code that passes an existing, weak test suite, or fail to create new, robust tests for their own logic. \textit{The Evidence:} The pack includes not just passing test logs, but the agent's test \textit{plan} and the new test cases it generated, allowing the reviewer to assess the verification strategy.

\textbf{(3) Exemplary SE Hygiene.} \textit{The Gap:} Agent-generated code can be functional but difficult to maintain, often violating project style guides or principles (e.g., DRY, SOLID). \textit{The Evidence:} The pack includes reports from static analysis, linting, and complexity checkers that help the reviewer assess whether the code is clean, readable, and free of avoidable technical debt.

\textbf{(4) Clear Rationale and Communication.} \textit{The Gap:} An agent's reasoning is often buried in low-level, verbose trajectory files or chat logs that are impractical for a human to audit. \textit{The Evidence:} The pack synthesizes this into a clear, human-readable summary (analogous to a PR description) explaining the approach and trade-offs, so the reviewer can evaluate the agent's decisions without reading raw logs.

\textbf{(5) Full Auditability.} \textit{The Gap:} True reproducibility is a major challenge due to agent non-determinism or environment changes. \textit{The Evidence:} The pack provides a ``frozen'' audit trail, including versioned links to the exact BriefingScript/MentorScript, tools, and agent trajectory used, supporting the auditing and re-execution of relevant checks.

To manage this density of information, the pack must support \textbf{``progressive disclosure,''} allowing a reviewer to see a high-level summary and then drill down into specific evidence like test logs or execution traces as needed. A concrete example of an MRP is provided in Appendix~\ref{sec:appendix_mrp}.

The same five criteria also guide the agent's self-evaluation. Before execution, the human coach defines the agent's objective in the BriefingScript and LoopScript by specifying the task's success criteria, hard constraints, required evidence, and resource limits. The agent then works to satisfy these requirements within the specified limits. Before submitting an MRP, the agent evaluates its work against the coach's requirements and the five criteria. Because unaided self-correction can be unreliable~\citep{huang2024selfcorrect}, the agent grounds these checks in external evidence, such as executed tests, static analysis, and reproducible traces, rather than in its own judgment alone~\citep{madaan2023selfrefine}. Passing these checks prepares the work for human review but does not establish correctness.

\subsection{The Tooling Gaps}

Finally, this motivational example shows a mismatch between the new agentic workflow and the tools that are available for both humans and agents:

\subsubsection{A New Workbench for the Human Developer: The ACE}\label{subsubsec:ace}
As Integrated Development Environments (IDEs) evolve from interactive code development toward agent orchestration, SASE proposes a workbench that connects execution, reusable guidance, and review evidence to support human coaching. We call this workbench the \textbf{Agent Command Environment (ACE)}. The ACE supports both 1-to-N collaboration, where one developer works with many agents, and N-to-N collaboration, where a human team coordinates a shared fleet of AI teammates.

Tools available in 2026 already support several of these functions. For example, Claude Code orchestrates subagents and background tasks~\citep{anthropic2025subagents, anthropic2025interactivemode}, Cursor runs multiple agents in parallel from its editor~\citep{cursor2025two}, and GitHub Copilot~\citep{github2025cloudagent}, Google Jules~\citep{google2025jules}, and OpenAI Codex~\citep{openai2025codex} operate cloud agents that open pull requests for asynchronous review. Cursor also supports version-controlled project guidance and team-wide rules~\citep{cursor2025rules}, and its cloud agents produce review artifacts such as logs and videos~\citep{cursor2025computeruse}. These capabilities provide a foundation for the ACE. SASE proposes to connect them through a common review process that compares candidate solutions against the same briefing criteria, links evidence to each candidate, tests revisions to guidance, and records specialist consultations as versioned decisions. The remaining gap is to integrate and evaluate these functions across a team's tasks, roles, and resource budgets.

This team-level setting requires multi-role governance, evidence-based review, and cross-functional consultation. Consultation Request Packs (CRPs) operationalize that collaboration: an agent can invoke a human specialist through a CRP, and the ACE routes, presents, and records the request. In this sense, the ACE treats humans as callable expertise endpoints while preserving the context needed for accountability. A concrete example of a CRP is provided in Appendix~\ref{sec:appendix_crp}.

The ACE should integrate the following capabilities around SASE artifacts and team workflows. It should support disciplined N-version programming, allowing a developer to visualize, compare, and mix components from multiple agent-generated solutions. It should also provide program-comprehension views that show the architectural impact of generated changes, moving beyond simple textual diffs. Because SASE relies on explicit artifacts, the ACE must support authoring, versioning, archival, and analysis for BriefingScript, MentorScript, and LoopScript. It must also help coaches curate the complex context agents need, which often differs from the context humans need. Finally, the ACE should support strategic agent management: coaches should be able to compose an agent team based on capability and cost, evaluate performance, and retrain, demote, or retire underperforming agents. The ACE must also let the coach ``jump in'' when direct implementation is more efficient than specification. For example, a developer should be able to switch into a traditional IDE view for a surgical code change, such as writing a complex mathematical formula, and then return to the coaching workflow.\footnote{The demo provides an example of performing a surgical code change when developing mobile apps: \url{https://www.linkedin.com/posts/ahmed-e-hassan_se3-harmonyos-ainativese-activity-7274452304278265856-7F7L}}

Versioning these artifacts also requires observability and revision control throughout the collaboration. Current tools already preserve parts of the development record. Claude Code supports conversation and code checkpoints, although these checkpoints do not capture every external effect of execution~\citep{anthropic2025checkpointing}, and pull-request workflows retain code changes, review comments, and CI results. SASE further calls for links among the version of a brief, the guidance available to an agent, its execution evidence, and the human decision that accepted or redirected the work. Retaining a conversation or reverting files alone does not establish these links or reconstruct the full execution state. Research must therefore identify which records and dependencies a team needs to audit a decision, reproduce a relevant check, or assess how a guidance change affected subsequent work.

Voice can serve as a complementary interaction modality for the ACE. It may be useful for high-level orchestration and mentorship tasks, especially when a developer needs to issue brief commands, dictate intent, or provide feedback without leaving the current work context. Research has shown that (a) speech can be faster than typing~\citep{gullberg2024scriptura}; (b) voice-assisted debugging can reduce context switching in software development workflows~\citep{amiri2025hearcodefailvoiceassisted}; and (c) current automatic speech recognition (ASR) systems, such as OpenAI's Whisper~\citep{whisper_2023}, make reliable transcription increasingly practical. The ASR layer need not perform reasoning; it can capture developer intent and pass the resulting text to downstream ACE tools and agents. Developer tools such as Talon Voice\footnote{\url{https://talonvoice.com}} and Cursorless for VS Code\footnote{\url{https://www.cursorless.org}} further demonstrate the feasibility of speech-based interaction with agents in accessibility and hands-free settings.

\subsubsection{An Optimized Workbench for the Agent: The AEE}\label{subsubsec:aee}
Just as humans need a new command center, agents require their own specialized environment designed for their unique capabilities. The tools that excel for humans are often suboptimal for agents. Much of modern SE has focused on creating high-level tools that reduce cognitive load for humans. This optimization, however, is often at odds with the needs of an agent. Agents have their own context, reliability, and resource constraints. Structured feedback and low-overhead operations can help them work within these constraints.

This necessitates the creation of an \textbf{Agent Execution Environment (AEE)}: a workbench built for agents, not humans. Instead of human-centric interfaces, the AEE must be equipped with agent-native tools that leverage their strengths. These tools might include hyper-debuggers capable of analyzing vast state spaces, powerful semantic search utilities, and structural editors that manipulate code as abstract symbolic structures rather than simple text. For example, early ideas considered structural manipulation through abstract syntax trees~\citep{poesia2022synchromesh}, while a more robust paradigm focuses on symbolic reasoning and formal environments~\citep{tu2026agenticverification}. The AEE also requires several additional capabilities. For example, because multiple agents may work on the same project concurrently, each agent should have an isolated, reproducible workspace with scoped credentials. Within this workspace, compilers, test runners, and profilers should provide diagnostics, test results, and runtime traces in machine-readable formats, enabling agents to act directly on failures rather than parse output intended for humans. The results of this work must then reach the humans who supervise it. Hence, the AEE needs a well-defined interface with the ACE, through which it receives the BriefingScripts, LoopScripts, and MentorScripts authored by humans and returns the CRPs and MRPs for human review. Beyond these task-oriented capabilities, the AEE must also include a robust monitoring infrastructure to manage the agents' operational health. This internal system would autonomously handle low-level issues such as spotting security vulnerabilities, flagging agents that are incurring unexpectedly high computational costs, or repairing and replacing broken virtual environments. The goal of this self-monitoring is to ensure that only significant problems requiring strategic human intervention are surfaced to the human in the ACE.

A shared IDE with agent panels, such as VS Code with GitHub Copilot~\citep{microsoft2025vscodecopilot}, offers an alternative that keeps familiar navigation and review workflows. SASE instead separates human supervision from agent execution so that each can be optimized independently. Google Antigravity's Manager and Editor views show a similar separation of orchestration from editing~\citep{google2025antigravity}. The ACE/AEE distinction also covers how artifacts pass between humans and agents and how execution infrastructure is provisioned.

\section{The Engineering Activities of SASE}\label{sec:sase_activities}

SASE is operationalized through structured engineering activities. This section does not present a definitive or exhaustive list. It offers an initial scaffold for organizing N-to-N collaboration among human coaches, AI teammates, and hybrid teams. The activities distinguish team-level agentic software engineering from solo agentic coding by making governance, evidence-based review, cross-functional consultation, and merge-readiness explicit. Consultation Request Packs (CRPs), for example, turn human consultation into a traceable team artifact. Teams could extend established approaches such as Scrum~\citep{schwaber2020scrum} and DevOps~\citep{kim2016devops} with agent roles, review gates, and machine-readable guidance, since these approaches already support coordination and automation. SASE makes explicit the responsibilities that such extensions leave implicit for agents. It specifies the teammate responsible for supplying context, the conditions under which an agent must seek guidance, and the evidence required at each handoff. Teams could adopt these responsibilities within their existing processes. The responsibilities matter more as agents take on longer tasks and work in parallel, because requirements, guidance, and decisions must survive more handoffs. SASE organizes established practices around them. We invite the community to challenge, refine, and extend these activities as Agentic SE matures.

\subsection{Briefing Engineering (BriefingEng): \texorpdfstring{Authoring BriefingScripts}{Authoring BriefingScripts}}\label{subsec:briefingeng}

In the Agentic SE era, the primary creative output of an engineer shifts from implementation logic to stating clear intent and guidance. Briefing Engineering (BriefingEng) is the activity that codifies this skill. It does not reinvent the wheel. It builds on decades of work in the Requirements Engineering (RE) and Agile/Scrum communities and adapts their principles to an agentic context. It is a hybrid discipline that combines requirements specification with architectural design, implementation advice, and test planning in a single artifact. As the ``brief'' becomes as important as the code itself, researchers and practitioners from these communities are well placed to lead the co-design of the next generation of specification practices for a future where humans guide and agents build.

\textbf{Purpose.} This activity moves beyond the common failure pattern of pasting a raw, vague ticket and expecting magic. It treats the mission brief as a first-class artifact, ensuring an autonomous agent receives a comprehensive and actionable work order. Unlike a traditional Software Requirements Specification (SRS), which is often implementation-agnostic, a BriefingScript is a specification for action: a version-controlled, testable, and machine-readable document that is as central to the engineering process as the source code.

\textbf{Actor.} The human Agent Coach.

\textbf{Workbenches.} All BriefingEng activities are centered in the Agent Command Environment (ACE). AI assistance can be used to help the coach author high-quality briefs by flagging ambiguity, surfacing edge cases, ensuring logical consistency, and generating property-based acceptance tests for the final output from the briefs.

\textbf{Artifacts.} The \textbf{BriefingScripts}. A BriefingScript is a structured document rather than a programming language. It can be serialized in any machine-readable format (e.g., Markdown, YAML, or JSON), allowing the specification to remain traceable and reviewable. In practice, a BriefingScript addresses five concerns: (1)~Goal \& Why, stating the objective and its business value; (2)~What \& Success Criteria, defining the scope through verifiable conditions and invariants; (3)~All Needed Context, curating the files, documentation, and known pitfalls the agent needs; (4)~Implementation Blueprint, providing strategic guidance and constraints rather than a low-level plan; and (5)~Validation Loop, specifying how the result will be tested and accepted. Appendix~\ref{sec:appendix_briefing} presents a concrete example. While its structure provides discipline, it is not intended to enforce a rigid, up-front specification. Instead, BriefingScripts are best authored through an interactive, iterative process, where the initial draft may be lightweight and refined over multiple cycles of agent interaction. This flexibility ensures briefs evolve naturally alongside the task, rather than locking humans and agents into an unrealistic ``waterfall'' style process. While the RE field has long pursued this goal, it becomes more feasible in the agentic context for three reasons:
\begin{enumerate}[leftmargin=*,nosep]
  \item The primary consumer is a machine, which requires and benefits from formal structure;
  \item Briefs are often for more granular tasks than a monolithic SRS, making formalization tractable;
  \item Modern AI assistants can help humans write these structured briefs, lowering the barrier to entry that hindered past formal methods.
\end{enumerate}

Alternative approaches include conversational prompting and formal specification. Conversational prompting supports exploration, but even a retained transcript requires explicit requirements and acceptance criteria to support review. Formal specifications offer stronger guarantees for properties that can be formalized, but require greater expertise and maintenance effort~\citep{woodcock2009formal}. BriefingScripts occupy a middle ground, providing reviewable task specifications that can incorporate formal constraints where appropriate. Structured specifications are well established, and recent tools such as Kiro~\citep{aws2025kirospecs}, which organizes requirements, design, and implementation tasks into documents, show that agents can help teams write and maintain them. BriefingEng further connects the specification to execution, evidence, and revision across a team.

\subsection{Agentic Loop Engineering (ALE): Disciplined Orchestration}

With a clear brief in place, Agentic Loop Engineering (ALE) governs how agents execute tasks. Just as BriefingEng builds on the work of RE and Agile, ALE is deeply rooted in the principles pioneered by the DevOps community. It transforms the agent's work from an opaque, black-box process into a disciplined, auditable, and reproducible workflow, moving beyond simple iterative cycles like the Plan-Do-Assess-Review (PDAR) loop. The declarative pipelines, infrastructure-as-code, and focus on observability that are central to modern DevOps are the direct precursors to the automated, agent-driven workflows defined by a LoopScript. We therefore see a crucial role for the DevOps community in designing the next evolution of CI/CD.

\textbf{Purpose.} This involves defining how agents work together (or alone), the patterns of their collaboration, and how they engage with their toolset. Such explicit direction is needed because agents cannot infer the ``stakes'' of a task. They may ``overthink'' a simple request or under-deliver on a critical one.

\textbf{Actors.} The workflow is orchestrated by the \textbf{human coach}, but the execution is performed by \textbf{agents}.

\textbf{Workbenches.} The workflow is defined in the \textbf{ACE}, but the agents execute it within the \textbf{AEE}.

\textbf{Artifacts.} The \textbf{LoopScripts}. Instead of relying on ad-hoc prompt hacking, the coach uses a declarative language like LoopScript to define the Standard Operating Procedure (SOP). Like all SASE artifacts, the LoopScript is a living document. A coach might dynamically adjust the workflow (for instance, by allocating more agents to a promising path or by adding a new review checkpoint if early results look uncertain), ensuring that the orchestration strategy is customized for a task when needed. A LoopScript can specify:
\begin{itemize}[leftmargin=*,nosep]
  \item Task Decomposition and Parallelization: A BriefingScript can be assigned to multiple agents, which could involve multiple instances of the same model or, more powerfully, a heterogeneous team composed of specialized agents. This reflects the emerging practice of using different models for their distinct strengths, for example, using a model like Gemini 2.5 Pro for high-level planning while using Claude Opus or Sonnet for detailed code generation. This makes powerful practices like N-version programming routine. For instance, a developer resolving seven tickets can trigger such a team to generate 28 distinct pull requests in parallel (4 per ticket), enabling creative exploration and increasing the probability of a successful outcome. The key metric shifts from single-task latency to overall system throughput.
  \item Workflow Strategy: The coach can define the required level of rigor, granting full autonomy for a simple bug fix while enforcing a strict, multi-stage review process for a critical security patch. The coach can also set per-task resource budgets, capping the wall-clock time, energy, or tokens an agent may consume before it must report back.
  \item Evidence-Based Acceptance Criteria: The LoopScript defines the structure of the final deliverable: a ``Merge-Readiness Pack.'' As detailed in Section~\ref{subsubsec:merge_pack}, this bundle proves the agent's work meets the five core criteria for being merged.
\end{itemize}

Emerging examples of LoopScript are discussed in Section~\ref{subsec:prompt}.

\subsection{AI Teammate Mentorship Engineering (ATME): Codifying Team Norms and Best Practices}

To ensure agent-generated code is not just functional but also maintainable and aligned with team culture, agent guidance must be treated as first-class code.

\textbf{Purpose.} To transform mentorship from an implicit, ephemeral activity (e.g., comments in a code review) into an explicit, evolving, and codified discipline. This is ``mentorship-as-code.''

\textbf{Actors.} Guidance is provided by the \textbf{human coach} and durably consumed by \textbf{agents}.

\textbf{Workbenches.} Mentorship rules are authored in the \textbf{ACE} and supplied to agents in the \textbf{AEE}.

\textbf{Artifacts.} The \textbf{MentorScripts}. These are structured, machine-readable rulebooks that codify project norms (aka tribal knowledge and aligned understandings). A MentorScript allows teams to define rules ranging from granular checks (``all new functions must have deterministic tests'') to high-level principles. This makes mentorship, once an implicit and ad-hoc activity, an explicit, reviewable, and continuously evolving discipline. MentorScript rules would be subject to their own quality gates, including linting, unit testing, and conflict detection. Such checks can catch ambiguity and conflicts, but whether a rule improves agent behavior is harder to establish, as Section~\ref{subsec:roadmap_atme} discusses. The AEE should record which rules were supplied to an agent so that a coach can investigate unexpected behavior. Tracing a decision back to the rules that influenced it is harder still, and prompt interpretation techniques such as PromptExp~\citep{dong2024promptexp} and reasoning observability techniques such as Watson~\citep{rombaut2025watson} offer starting points. This explicit guidance aims to reduce the burden on the human coach to infer complex rule interactions, although its effects on agent predictability and reliability remain to be evaluated. Fine-tuning offers another way to teach agents team conventions and can complement explicit guidance. However, MentorScripts keep those conventions in a form the team can read and edit. Changing a fine-tuned behavior requires retraining and re-evaluating the model, and the result may not carry over to a different base model. Agent Skills~\citep{anthropic2025skills} now package instructions and supporting resources in a portable, open format. A skill implements part of what a MentorScript specifies, namely reusable guidance that an agent loads on demand. ATME further proposes behavioral tests, conflict checks, and traceability from guidance to review outcomes.

\textbf{Structured Mentorship.} The review process generates a critical, multidimensional artifact. When a coach provides feedback, it is captured systematically across the explicit, inferred, and multi-solution dimensions previously outlined in Section~\ref{subsubsec:mentorship}.

Section~\ref{subsec:prompt} discusses meta-prompt files (e.g., CLAUDE.md, AGENTS.md) as early examples of MentorScript-like artifacts.

\subsection{Agentic Guidance Engineering (AGE): Leveraging the Human in the Loop}\label{subsec:age}

While BriefingEng initiates work, \textbf{Agentic Guidance Engineering (AGE)} governs the structured role of the human in reviewing and responding to agent-generated artifacts and clarification requests (e.g., during the BriefingEng and ATME activities). AGE elevates the human from a passive approver of outputs to an active, on-demand consultant who intervenes precisely where their expertise adds the greatest value.

\textbf{Purpose.} To formalize and optimize human participation in the agentic loop, ensuring that when agents escalate issues through a Consultation Request Pack (CRP) or submit a Merge-Readiness Pack (MRP), human input is efficient, targeted, and becomes a durable part of the project record.

\textbf{Actors.} The human engineer (who may be the task initiator or a domain specialist).

\textbf{Workbenches.} All AGE activities are performed within the \textbf{ACE}, which provides an inbox-like interface for triaging CRPs, auditing MRPs, and issuing structured resolutions.

\textbf{Artifacts.} AGE consumes two types of agent-generated artifacts: (a) \textbf{Consultation Request Packs (CRPs)} and (b) \textbf{Merge-Readiness Packs (MRPs)}, and produces \textbf{Version Controlled Resolutions (VCRs)}.
CRPs are generated when an agent requires human input to proceed. A CRP is contextualized by the active BriefingScript, potentially triggered by LoopScript or MentorScript rules, and documents the specific uncertainty or decision point. MRPs are structured evidence bundles submitted for human approval, designed to demonstrate functional completeness, sound verification, engineering hygiene, rationale, and auditability. The outcome of AGE activities is a Version-Controlled Resolution (VCR). Each resolution is explicitly linked to the artifact that it addresses (CRP or MRP), preserving traceability and enabling downstream auditing and learning.

AGE must also decide when an agent should request human involvement. In SASE, explicit rules govern this decision instead of leaving it entirely to the agent's own discretion. An agent issues a CRP in four situations: (1) a LoopScript or MentorScript rule requires consultation, e.g., before a security-sensitive change; (2) the agent's self-evaluation (Section~\ref{subsubsec:merge_pack}) identifies uncertainty or a trade-off that cannot be resolved using the available context; (3) repeated validation failures suggest that further autonomous attempts are unlikely to succeed; (4) the agent exhausts the resource budget set in its LoopScript. A coach may also intervene without a CRP. Signals that can prompt such intervention include validation results that diverge across attempts, resource consumption that outpaces the budget, repeated CRPs on the same task, and outputs that drift from the BriefingScript.

\subsection{AI Teammate Lifecycle \& Infrastructure Engineering (ATLE \& ATIE): Building the SE for Agents Foundation}

To realize the full potential of Agentic SE, we must simultaneously support the SE activities of agents (i.e., SE for Agents). This involves fundamentally rethinking our tools, practices, and artifacts when the primary actor is an agent, not a human. The best practices that have served us for decades, designed around the constraints of human cognition and patience, must now be re-examined and, in many cases, inverted. Engineering this new, agent-centric foundation is a monumental task that falls squarely within the expertise of the Platform Engineering community. As agents begin to operate at scale, the need for the robust, secure, and scalable platforms that are the core mission of Platform Engineering becomes paramount. Their work is essential in building the next generation of internal developer platforms, not for humans, but for the fleets of autonomous agents that will inhabit them, ensuring these systems are reliable, efficient, and secure.

\textbf{Purpose.} To engineer the agent's environment (the AEE), enable agents to retain memory and learn over time (ATLE), and build the agent-native toolchains they need to operate effectively (ATIE).

\textbf{Actors.} The fundamental shift is in the actor itself. For decades, SE has been optimized for the human developer. In an agent-centric world, the primary actor is \textbf{computational}. The human engineer's role evolves into that of a strategist, mentor, and conductor, serving as the ultimate arbiter of value and the indispensable conduit for tacit, ``tribal'' knowledge. The ultimate goal is not agents that are perfect out of the box but ones that can learn and ramp up first.

\textbf{Workbenches.} This work is fundamentally about architecting the \textbf{Agent Execution Environment (AEE)}.

\textbf{Core Concepts.} ATLE and ATIE are grounded in several foundational concepts that collectively define the software engineering infrastructure required for agent-centric development. These concepts span memory, coding practices, tooling, team organization, and long-term agent evolution.

\textbf{(1) Persistent Memory (ATLE):} Agents are equipped with long-term memory of project history and decision logs, allowing them to maintain continuity across tasks without the coach repeatedly supplying the same guidance.

\textbf{(2) Agent-First Code Practices (ATLE):} With the agent as the primary actor, long-standing process principles must be re-evaluated. For instance, the \textbf{``Don't Repeat Yourself'' (DRY)} principle is often reversed. Code cloning, a source of maintenance debt for humans, can become a viable strategy for an agent, as it simplifies its reasoning process while the downside (updating all instances) is trivial. This theoretical shift is supported by industry observations, such as the GitClear report noting sharp increases in code duplication on GitHub since the emergence of Copilot.
Furthermore, the \textbf{ROI of Clean Code} (high cohesion, low coupling, comprehensive documentation) becomes clear, as these practices make a codebase a more fertile environment for agents to inhabit. Such efforts are analogous to making an open-source project accessible to novices through a well-defined plugin architecture, in turn attracting and growing a large community around that project. Previously, this meant redirecting resources away from immediate business needs like feature development. Now, a small human team, aided by agents, can perform this cleanup, which in turn unlocks massive productivity gains for its fleet of feature-developing agents.
This new model also favors programming languages with strong, compile-time safety guarantees, such as \textit{Rust and TypeScript}. The up-front effort to satisfy a strict compiler is less of a barrier for an agent, and the payoff is immense, as the strong type system prevents entire classes of bugs by construction.

\textbf{(3) Agent-Native Toolchain (ATIE):} The tools built for human developers are often ill-suited for agents. We have spent decades building tools like IDEs and visual debuggers to reduce cognitive overload, but an agent has no such limitations. This shifts the entire optimization landscape. SE for Humans has historically focused on precision@K for a small K, because human time is precious. In an agent-first world, precision@100 is perfectly acceptable if a subordinate agent can post-process the results, opening up entirely new avenues for automated analysis where the human is completely out of the loop. However, agent time is not free. Each retry and each subordinate agent consumes wall-clock time, energy, and tokens or compute resources. A larger K is therefore beneficial only when the resulting improvement outweighs the cost of processing the additional results. This trade-off depends on the task, model, and deployment, so orchestration must budget agent effort as carefully as human attention.

Expressive feedback is paramount. Rust's toolchain exemplifies this, as its rich, constructive compiler messages enable agents to learn quickly from failures, providing a blueprint for agent-friendly environments. The path forward involves creating Agent-Native Model Context Protocol~(MCP) servers~\citep{hasan2025mcpserver} that return deep, interpretable feedback and support agent-driven refinement of tool usage descriptions. This kind of self-improving tooling loop, already being adopted manually by teams like Anthropic by optimizing their MCP descriptions to suit agents instead of humans, is proving essential for making agentic SE robust, scalable, and fault-tolerant.

\textbf{(4) Engineered Multi-Agent Teams:} While some human-centric rules are inverted, foundational principles for managing system complexity, like Modularity and Separation of Concerns, become even more critical. This is driving a key trend away from monolithic, general-purpose agents and toward engineered multi-agent teams. These systems are structured workflows where agents are assigned specialized roles like ``planner,'' ``coder,'' ``tester,'' and even a ``critic'' to create an internal feedback loop.

While this specialization is a practical solution to the core limitations of today's large models (protecting each agent's context from the ``pollution'' of side-tasks and an overwhelming number of tool choices), it also points to a more robust, long-term architectural principle. A modular design improves the interpretability of the system's behavior, as the trajectory of a specialized agent is far easier to audit than the interleaved reasoning of a monolithic one. This also paves the way for an ecosystem with highly specialized agents. This, in turn, could lead to the emergence of ``agent stores'': digital marketplaces analogous to today's app stores. In such a future, a bespoke team could be dynamically composed, with a coordinating agent (whether human or AI) selecting best-in-class agents from various vendors for specific roles, such as a ``React Refactoring Agent'' or a ``Python Security Audit Agent.'' Moreover, Multi-Agent Teams provide critical security and control benefits. By assigning specific tools as well as security and resource-usage policies to specialized agents (e.g., a ``Test-Agent'' that can run tests but not commit code), the system limits the potential ``blast radius'' of a misbehaving or compromised agent.

These observations are already manifesting in different ways. For instance, the evolution of commercial Autonomous Agents like Anthropic's Claude Code has shifted from a monolithic agent to a multi-agent architecture where specialized sub-agents are spawned for specific tasks. A more comprehensive example is the BMAD (Breakthrough Method for Agile AI-Driven Development) framework, which takes the ``team'' metaphor literally, organizing agents into a full-fledged agile structure to tackle complex projects.

\textbf{(5) Lifetime Teammates:} The largest shift may be the move from stateless agents to persistent teammates that learn and grow over time. This is the focus of \textbf{AI Teammate Lifecycle Engineering (ATLE)}, a discipline aimed at giving agents memory and long-term context.

\begin{itemize}[leftmargin=*,nosep]
  \item \textbf{From Contractors to Partners:} The goal is to evolve agents from ``one-off contractors'' who start every new task from scratch to ``lifelong partners'' who retain institutional knowledge. Early examples of this concept, like the \textbf{DeepWiki} used by the Devin agent, allow an agent to build and refer to its own documentation and decision logs across multiple tasks, creating continuity and preventing it from repeating mistakes.
  \item \textbf{Proactive Maintenance:} An agent with persistent memory and access to the codebase can become a proactive partner. During idle compute cycles, the ACE can schedule agents to perform valuable maintenance tasks, such as scanning for technical debt, identifying documentation gaps, or proposing code refactorings. These proposals would be filed as new BriefingScripts, entering the standard SASE workflow for human review and prioritization, thus transforming maintenance from a reactive chore into a continuous, autonomous improvement process and eventually moving SE 3.0 into SE 4.0, where we remove the requirement to obtain human approval for such optimization activities.
\end{itemize}

A comprehensive industry example of these principles is the BMAD framework, discussed in Section~\ref{subsec:multiagent}.

\section{Discussion}\label{sec:discussion}

\subsection{Embracing the Bitter Lesson in the Agentic SE Era}

At first glance, our emphasis on structured processes (from BriefingScripts to structured orchestration via LoopScripts) might seem to run counter to the core message of Rich Sutton's \textbf{``Bitter Lesson''}~\citep{sutton2019bitter}. That lesson powerfully argues that general methods that scale with computation and data ultimately triumph over approaches that rely on baking in specific human knowledge and processes. One might therefore think that our attempts to structure Agentic SE are a futile effort to impose human-centric designs on this new era.

However, the lesson applies most strongly where data is abundant, such as at lower levels of abstraction or for common problems like building a web application. It applies less cleanly to novel tasks or niche domains where training data is scarce. In these settings, relying on large-scale data alone is inefficient, and a human is still needed to provide the overall structure and connect the dots.
This need comes from constraints on both current and future AI. AI systems lack the embodied human experience necessary for physical-world verification, contextual judgment, and deep ethical reasoning~\citep{AISwiss2025}.
SASE therefore does more than encode human knowledge. It builds human-AI partnerships around complementary roles, in which humans provide strategic and ethical guardrails with enough precision for AI agents to act.
Even within a novel software system, there may be repetitive sub-problems where an agent can and should be granted full autonomy. We also note that in any system, a baseline of structured interactions, a Standard Operating Procedure (SOP), is essential for coordination and governance, especially in regulated settings.

Therefore, our vision does not reject scale. It seeks to create the conditions under which scale succeeds reliably and reproducibly in a complex engineering context. Our focus also extends beyond process control to traceability, human-agent communication~\citep{wu2025humanevalcomm}, and an agent-first mindset in how we structure codebases and design tools. SE, especially in this new era, is as much about the journey (the collaboration, the evolution of intent, the trail of decisions) as it is about delivering the final outcome.

We believe software engineers must embrace the Bitter Lesson as a guiding principle. The core skill of the modern super software engineer is mastering the duality of control: strategically deciding when to impose a structured workflow (for novel, high-level tasks) and when to ``let the agent loose'' on a well-defined problem where scaled learning pays off. Managing a team of brilliant experts is similar. One must know when to provide a detailed work order and when to trust a team member with full autonomy.

\subsection{Agentic Software Engineering, Not Just Agentic Coding: The Centrality of N-to-N Collaboration}

A foundational premise of this paper is the distinction between agentic coding and agentic software engineering. Agentic coding augments a solo activity to boost individual productivity.

Software Engineering (SE), by contrast, has always been a \textbf{team sport}. SE involves not only producing code but also managing complexity, coordinating across diverse roles, reconciling competing stakeholder needs, and ensuring the long-term sustainability of shared artifacts. These collective challenges call for faster structured collaboration, not only faster individuals.

Structured Agentic Software Engineering (SASE) is explicitly designed for this broader scope. It provides the artifacts, processes, and workbenches necessary to support \textbf{N-to-N collaboration}, where many humans and many agents interact as a coordinated team. This model appears at several levels:
\begin{itemize}[leftmargin=*,nosep]
    \item \textbf{1-to-N Human-Agent Collaboration:} A single human orchestrates and mentors a fleet of agents, directing parallelized workstreams.
    \item \textbf{1-to-N Agent-Human Collaboration:} An agent escalates a Consultation Request Pack (CRP) to the appropriate human specialist, enabling targeted, domain-specific feedback.
    \item \textbf{N-to-N Hybrid Collaboration:} Multiple humans collectively oversee and mentor a shared pool of agents, while agents collaborate with one another or even with specialized sub-agents.
\end{itemize}

For instance, an agent may route a database schema issue to the designated database architect, who provides feedback through the ACE. In more advanced settings, that ``architect'' role may itself be filled by another specialized agent. Without explicit, durable artifacts such as the CRP, these complex multi-actor workflows would be ephemeral, untraceable, and unmanageable. SASE requires these interactions to leave structured, auditable records, which turns ad-hoc agentic coding into a disciplined engineering practice.
\section{From Vision to Reality: A Research Roadmap for Structured Agentic SE and Its Implications}\label{sec:roadmap}

\begin{figure}[t]
    \centering
    \includegraphics[width=0.9\textwidth]{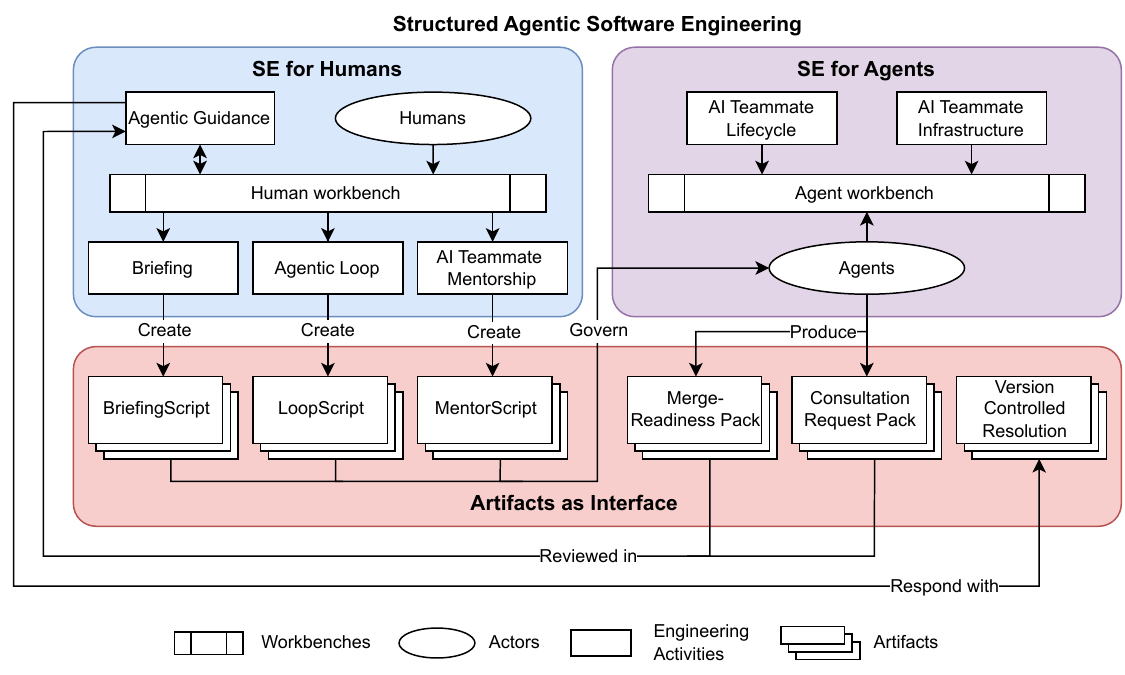}
    \caption{The Structured Agentic Software Engineering~(SASE) framework: dual domains for humans and agents, engineering activities, and artifacts.}
    \label{fig:sase_roadmap}
    \Description{SASE organizes human and agent workbenches around shared artifacts. Briefing Engineering, Agentic Loop Engineering, and AI Teammate Mentorship Engineering create BriefingScripts, LoopScripts, and MentorScripts. AI Teammate Lifecycle and Infrastructure Engineering support agents. Agents produce Merge-Readiness Packs and Consultation Request Packs, which humans review through Agentic Guidance Engineering and answer with Version-Controlled Resolutions.}
\end{figure}

The research directions outlined below are not meant to be comprehensive. They aim to shift the community's focus toward the new challenges of engineering with and for intelligent agents. Fig.~\ref{fig:sase_roadmap} outlines the relations between the tools, activities/processes, actors, and artifacts across the two modalities. For each direction, we also identify the specific challenges involved.

\subsection{Briefing Engineering (BriefingEng)}

\textbf{Formalizing BriefingScript.} Research is needed on languages and schemas that express goals, constraints, invariants, domain context, and acceptance criteria without forcing premature design decisions. The central challenge is balancing expressive power, learnability, and machine-checkable structure. A brief written largely in natural language cannot be mechanically checked for consistency or completeness, and the more structure a schema imposes to enable such checks, the more design decisions it forces the coach to make up front.

\smallskip\noindent\textbf{AI-Powered Authoring and Review.} Briefing tools should help coaches detect ambiguity, surface missing context, generate edge cases, and check consistency across clauses. One challenge is enabling such assistants to guide engineers toward property-based acceptance criteria rather than brittle examples. Another challenge is that the assistant is itself a model and may share the blind spots of the agent it reviews~\citep{panickssery2024selfpreference}.

\smallskip\noindent\textbf{Traceability and Multi-Solution Review.} Trustworthy briefs require traceability from generated code and evidence back to the briefing clauses that motivated them. This need becomes sharper under N-version programming, where coaches must compare, combine, and justify choices across multiple agent-generated alternatives. Cognitive observability tools such as Watson~\citep{rombaut2025watson} record an agent's reasoning and offer a starting point, but SASE requires observability across interacting artifacts, agents, and review decisions. A further challenge is that an agent's stated reasons may not reflect its actual reasoning~\citep{turpin2023unfaithful}.

\subsection{Agentic Loop Engineering (ALE)}\label{subsec:roadmap_ale}

\textbf{Designing LoopScript.} A declarative LoopScript language should capture task decomposition, parallel execution, review checkpoints, escalation rules, and evidence requirements. Research can build on business-process and DevOps automation, but SE-specific workflows need stronger links to code, tests, risks, and human review. The analogy to DevOps automation also has limits. A CI/CD pipeline typically fixes its task graph before execution, while an agent workflow creates subtasks at runtime and may revert earlier work. Hence, termination and resource bounds cannot be checked statically, and progress is difficult to measure.

\smallskip\noindent\textbf{Human-in-the-Loop Control.} Agents will explore unproductive paths, misread constraints, or need domain judgment. We need interaction mechanisms that let a coach pause a workflow, redirect a branch, add context, or stop low-value work without restarting the whole loop. However, the coach has no reliable signal for when to intervene, since an agent that is stuck and one that is exploring may appear similar from the outside.

\smallskip\noindent\textbf{Evidence Packs and Feedback Signals.} Merge-Readiness Packs require standards for sufficient evidence of correctness, security, quality, and rationale. Tool feedback also needs to become more informative for agents. A structured compiler diagnostic, test failure, or static-analysis finding can guide the search far better than an opaque failure signal. Because the agent assembles the evidence used to certify its own work, the review is no longer fully independent. In addition, a diagnostic precise enough to guide repair may also enable the agent to pass the check without fixing the underlying fault.

\subsection{AI Teammate Mentorship Engineering (ATME)}\label{subsec:roadmap_atme}

\textbf{Developing MentorScript.} MentorScript needs abstractions for rules that range from concrete style constraints to architectural principles. The language must be expressive enough for nuanced guidance, but simple enough for teams to review as ordinary engineering artifacts. However, many norms are tacit, and experts recognize a violation without being able to specify the rule's boundaries in advance. Hence, precise rules become brittle, while general rules leave interpretation to the model.

\smallskip\noindent\textbf{Quality Assurance for Mentorship Rules.} If mentorship becomes code, it needs linting, testing, conflict detection, and regression checks. Researchers should study how to verify that new rules improve agent behavior without causing unintended changes elsewhere. Practitioners already maintain such guidance as code. In Claude Code plugin marketplaces, skill instruction files frequently change together with the scripts they invoke, and most of these co-changes are functionally coupled~\citep{hereiz2026maintenancecoevolutionagentplugins}. However, rules are harder to verify than program code because their meaning depends on the model interpreting them. The same rule may be followed by one model and ignored by another (or even by a later version of the same model). Furthermore, conflicts between rules may emerge only in the agent's behavior on particular inputs, making them difficult to detect through inspection alone.

\smallskip\noindent\textbf{Learning and Explaining Team Norms.} Agents may infer candidate MentorScript rules from repeated human feedback, but those rules should remain reviewable and auditable. Developers also need explanations that connect an agent's decision to the specific rules considered, drawing on prompt interpretation and reasoning-observability techniques such as PromptExp~\citep{dong2024promptexp} and Watson~\citep{rombaut2025watson}. However, a correction rarely indicates whether it reflects a team norm or a passing preference, so rules learned from corrections risk fossilizing idiosyncrasies. Also, an agent's explanation is generated text rather than a trace of its computation, so it may be plausible but unfaithful~\citep{turpin2023unfaithful}.

\subsection{AI Teammate Lifecycle Engineering (ATLE)}

\textbf{Models for Persistent Agent Memory.} Agents need mechanisms for retaining project history, decision logs, architectural rationale, and lessons from review. This includes both continual-learning methods~\citep{shi2024continual,guo2025comprehensive} and external memory structures such as graphs, vector stores, and decision records. Since unbounded history can overflow context windows or introduce irrelevant evidence, SE-specific memory compression must preserve semantic details in code, builds, tests, and dependencies~\citep{wingate2022prompt, fei2024extending, chevalier2023adapting, xu2024soft}. The machine learning~(ML) and natural language processing~(NLP) communities already provide building blocks such as virtual context management modeled on operating systems~\citep{packer2023memgpt}, memory that reorganizes itself as new notes arrive~\citep{xu2025amem}, memory layers and temporal knowledge graphs built for production use~\citep{chhikara2025mem0, rasmussen2025zep}, hierarchical working memory for long tasks~\citep{hu2025hiagent}, and compression that keeps a growing memory bank tractable during continual adaptation~\citep{katraouras2026memorybank}. Two recent surveys map this design space~\citep{zhang2025memorysurvey, mei2025contextsurvey}. Document engineering adds a complementary source of context by converting complex documents into structured representations that support context curation~\citep{livathinos2025docling}, and layout-aware retrieval can ground an agent's answers in specific regions of those documents~\citep{brown2026doclayout}. Memory must preserve hard-won lessons without retaining outdated conventions. It also introduces hidden state. Two agents with the same model and brief may behave differently because of what they remember, and an incorrect (or stale) memory may cause more harm than benefit.

\smallskip\noindent\textbf{Proactive Maintenance.} Persistent agents can scan for technical debt, documentation gaps, fragile tests, or refactoring opportunities during idle cycles. Research should study how to schedule such work, estimate its value, and present proposals without distracting teams from higher-priority development. The difficulty is that the value of such work is hard to estimate, because the cost of technical debt materializes only when the affected code must change. In addition, every proposal consumes reviewer attention, so an agent that proposes changes too freely may add more review burden than it removes.

\smallskip\noindent\textbf{Economics of Agent-First Code.} Agentic SE may change the cost model behind long-standing principles. For example, duplication may be easier for agents to update consistently, while strong type systems may become even more valuable because compiler feedback helps agents repair mistakes. Replacing such anecdotes with empirical evidence faces two challenges. First, model capabilities may change faster than a longitudinal study can be completed, making it difficult to separate the effect of a practice from that of a model upgrade. Second, maintainability and readability have traditionally been defined relative to human cognitive limits and may need to be redefined if agents become the primary maintainers.

\subsection{AI Teammate Infrastructure Engineering (ATIE)}

\textbf{The Post-IDE Human-Agent Interface.} If agents perform much of the direct editing, the human-facing environment becomes a command center for specifying intent, comparing alternatives, routing consultations, and auditing evidence. Research should move toward interfaces for orchestration, review, and structured mentorship. The ironies of automation~\citep{bainbridge1983ironies} also apply here. A coach who no longer edits the code loses the familiarity that makes review judgments reliable, but must evaluate more work across multiple concurrent workstreams.

\smallskip\noindent\textbf{Distributed Compute Fabrics for Agents.} Multi-agent SE needs runtime support for isolation, reproducibility, scheduling, and cost control. The declarative nature of LoopScript can expose workflow structure for optimization, connecting SASE to related work on SLA-aware and context-aware CodeLLM serving~\citep{kishan2025contextaware}. Reproducibility is difficult to guarantee because model outputs can vary even under fixed settings (e.g., zero temperature). Isolation is also more challenging than for conventional jobs. For example, an agent runs code that it wrote and that no one has reviewed yet, but it needs access to repositories and package registries to make progress. Hence, the agent can be neither fully sandboxed nor fully trusted.

\smallskip\noindent\textbf{Agent-Native Toolchains.} Today's tools expose interfaces optimized for humans, so agents often fall back to brittle textual search and shell commands. Future toolchains should provide machine-readable protocols, structured diagnostics, semantic search, and self-describing operations that agents can use and improve over time. Protocols such as MCP, along with skills and plugins, already standardize how agents discover and invoke tools, but they do not necessarily change what those tools return. Many MCP servers wrap existing human-facing tools and expose the same unstructured text through a new channel, while their tool descriptions are often too vague or verbose for agents to use efficiently~\citep{hasan2026mcpsmell}.

\subsection{The Human Differentiator: A Call to Reimagine SE Education}

The SE field is defined by its four pillars: actors, processes, tools, and artifacts. While the SASE vision provides a structured approach for the latter three, we must not forget that the actor (the human engineer) remains the most critical factor. The 100x and 1,000x productivity gains emerging today are not the result of magical tools, but of skilled individuals who have mastered the art of working with agents. They demonstrate that even with today's nascent technology, the human's ability to specify intent and provide strategic oversight is the ultimate differentiator.

Even if the entire SASE framework were realized tomorrow, it would not automatically create a generation of 1,000x engineers. It would provide the scaffolding, but the ability to use that scaffolding effectively (to excel at Briefing Engineering, to design elegant LoopScripts, to codify insightful mentorship) will still separate the exceptional from the average. The human is not being automated away, but being elevated from a crafter of code to a conductor of agents.

This raises a critical challenge that extends beyond research: we must fundamentally rethink SE education. Current curricula are largely designed to train students to \textit{be} the agent (to write code, create tests, and handle low-level implementation). In the agentic era, we must instead train them to \textit{manage} fleets of agents. This requires a profound pedagogical shift away from pure implementation and toward strategic skills: system-level thinking, architectural reasoning, rigorous specification, and the art of mentorship-as-code. Rather than simply adding a ``prompt engineering'' module to an existing course, this is a call for a deep and holistic reimagining of what it means to educate the next generation of software engineers. The difficulty is that students have traditionally built the judgment needed to review code by writing it themselves, so a curriculum that moves effort away from implementation must find other ways to develop that judgment. In addition, agent capabilities change faster than curricula, so the specific skills taught today may soon be automated. This challenge is further compounded by the concurrent shift in the nature of software itself toward ``FMware'' (FM-powered applications), a separate but equally transformative trend that also demands its own educational rethinking, though we do not address that here.

\section{Related Efforts in Agentic Software Engineering}\label{sec:related_efforts}


\subsection{Iterative and Prompt-Driven Workflows}\label{subsec:prompt}

\textbf{PDAR with Product Requirement Prompts (PRPs).} The Plan-Do-Assess-Review loop formalizes a single task's lifecycle. A human and AI plan, a dev-agent implements, an agent self-assesses, and a human reviews. PRPs act as the ``minimum viable packet'' capturing goals, justification, acceptance criteria, and curated context. This aligns with SASE's emphasis on structured, testable intent. PDAR, however, is scoped to one-off execution. It does not by itself establish durable mentorship, agent lifecycle learning, or cross-task traceability that SASE treats as first-class.

\smallskip\noindent\textbf{Agent Skills and Plugins.} Superpowers~\citep{vincent2025superpowers} encapsulates best-practice prompts into predefined skills/plugins for agents to perform tasks like brainstorming, reviewing code, and running debugging workflows. Versioned, shareable skills can deliver the guidance that MentorScripts capture, and ATME adds regression tests, conflict checks, and traceability to review outcomes.

\smallskip\noindent\textbf{Meta-prompt files (AGENTS.md, CLAUDE.md).} Practitioners use project-level configuration files (e.g., CLAUDE.md, .clinerules, AGENTS.md) to load ``institutional knowledge'' before every agent task, codifying style guides, architectural constraints, and lessons learned into a continuously improving ``employee handbook'' for AI teammates. This grassroots practice leaves an open question. The community has no consensus on what such files should contain or how detailed they should be, so future work must define languages like MentorScript and also discover best practices for using them.

\subsection{\texorpdfstring{Agent and Multi-Agent Frameworks}{Agent and Multi-Agent Frameworks}}\label{subsec:multiagent}

Research prototypes such as ChatDev~\citep{qian2024chatdev} and MetaGPT~\citep{metagpt2024} assign agents software development roles, such as designer, programmer, and tester, and coordinate their collaboration through structured conversations or standardized operating procedures. BMAD~\citep{bmad2025method} brings this idea into practice and organizes agents into agile roles (e.g., Product Owner, Architect, Developer, Tester). Up-front agentic planning yields PRDs and designs, a Scrum-like shard step creates ``story files'' with focused context, and specialized agents execute in parallel. BMAD's strengths (role specialization, task sharding, and high parallelism) map well to SASE's N-version programming and orchestration. SASE goes further by (i) converting review feedback into persistent MentorScript rules (mentorship-as-code), and (ii) specifying the environments and disciplines (ACE for human coaching and orchestration, AEE for agent execution, and ATLE/ATIE for memory, lifecycle, and agent-native tooling). Meanwhile, coding agents such as SWE-agent~\citep{yang2024sweagent}, OpenHands~\citep{wang2024openhands}, and AutoCodeRover~\citep{zhang2024autocoderover} provide purpose-built interfaces for navigating code, editing files, and running tests in real repositories. These agents can serve as executors in the AEE. SASE complements them with a BriefingScript that defines their tasks and an MRP that provides humans with the evidence needed to make merge decisions.

\subsection{Future of Software Engineering}

Recent papers describe a shift from code-completion tools toward socio-technical systems in which developers specify intent, coordinate agents, evaluate evidence, and remain accountable for outcomes. \citet{Ahmed2025JourneyAISE} survey the AI for SE landscape and identify trust in AI-generated code and human-AI collaboration as key open challenges. \citet{Gao2025ChallengesLLM4SE} emphasize hallucinations, limited contextual understanding, and shortcomings in current evaluation methods. \citet{Terragni2025FutureAISE} and \citet{Qiu2025FromTodaysCode} extend this trajectory to full-lifecycle AI-driven development and the changing daily routine of developers. \citet{He2025MultiAgentSE} focus on role-specialized multi-agent systems, while \citet{Grundy2025SE4Humans} emphasize developer experience, cognitive load, and human accountability.

\subsection{How SASE Aligns and Differentiates}

\textbf{Alignment.} SASE adopts the best of these efforts, such as PRP-style briefs for intent formalization, PDAR-style iterative loops, and BMAD-like multi-agent parallelism.

\smallskip\noindent\textbf{Differentiation.} SASE combines these pieces into an engineering methodology with five distinguishing features:

\begin{enumerate}[leftmargin=*,nosep]
    \item \textbf{Mentorship-as-Code (ATME).} Review guidance becomes a version-controlled, testable MentorScript, enabling cumulative, auditable improvement across tasks and teams. Examples include Cursor's project rules~\citep{cursor2025rules} and Anthropic's Agent Skills~\citep{anthropic2025skills}, which store reusable guidance as files that teams can version-control and agents can load across tasks.
    \item \textbf{Dual Workbenches.} The Agent Command Environment (ACE) optimizes human cognition for specification, orchestration, and evidence-based review, and the Agent Execution Environment (AEE) optimizes for agent strengths (e.g., massive parallelism). An example is Google Antigravity~\citep{google2025antigravity}, which separates a Manager view for orchestrating agents from an Editor view for hands-on coding.
    \item \textbf{Merge-Readiness as the Target Artifact.} The loop's output is a Merge-Readiness Pack, which is a progressive-disclosure bundle proving functional completeness, sound verification, SE hygiene, rationale, and full auditability. An example implementation is GitHub Agentic Workflows~\citep{github2026agenticworkflows}, which can trigger agents through GitHub Actions and generate artifacts for a pull request while leaving the merge decision to a human.
    \item \textbf{Consultability as a First-Class Artifact.} SASE introduces the Consultation Request Pack as a structured artifact for agent-initiated human consultation. This elevates humans to callable experts and enables traceable cross-role handovers, shifting the paradigm from solo agentic coding to team-based Agentic Software Engineering. Example implementations include agents that ask questions through pull request comments~\citep{github2025copilotprcomments} or in team chat~\citep{cognition2025devinslack}.
    \item \textbf{Lifecycle \& Infrastructure (ATLE \& ATIE).} Agents become persistent teammates with memory, observability, and secure, hermetic execution, shifting from stateless contractors to evolving collaborators. Example implementations include the cross-session memory capabilities of OpenAI Codex~\citep{openai2026codexmemories} and Claude Code~\citep{anthropic2025memory}.
\end{enumerate}

\section{Conclusion}\label{sec:conclusion}

Agentic Software Engineering (SE 3.0) demands more than incremental adjustments to existing SE practices. In this paper, we presented Structured Agentic Software Engineering (SASE), a conceptual framework for making agentic SE more structured, predictable, and trustworthy. Our core contribution is a duality between SE for Humans and SE for Agents, which reimagines the field's actors, processes, tools, and artifacts around human-agent collaboration.

SASE operationalizes this duality through dedicated environments (ACE and AEE) and version-controlled artifacts (BriefingScript, LoopScript, MentorScript, CRP, and MRP). These mechanisms shift the human role from direct implementer to strategic ``Agent Coach and Orchestrator,'' with important implications for SE education.

We offer SASE not as a final solution, but as a scaffold for community discussion and refinement. By building disciplined foundations for human-agent collaboration, the SE community can move beyond impressive but brittle demonstrations. The future of SE will be defined not by agent speed alone, but by our ability to mentor, orchestrate, and trust agents as engineering partners.

\bibliographystyle{ACM-Reference-Format}
\bibliography{main}

\newpage
\appendix
\section{Examples of SASE Artifacts}\label{sec:appendix_examples}

To make the SASE framework more concrete, this appendix provides illustrative examples of three key artifacts that span the complete task lifecycle: (1) a BriefingScript that initiates work, (2) a Consultation Request Pack (CRP) generated during execution when human guidance is required, and (3) a Merge-Readiness Pack (MRP) submitted upon task completion. Together, these examples illustrate the structured dialogue between the Agent Command Environment (ACE) and the Agent Execution Environment (AEE).

\subsection{BriefingScript Example}\label{sec:appendix_briefing}

Listing~\ref{lst:briefingscript} shows a BriefingScript for implementing rate limiting on a REST API. The example addresses the five concerns defined in Section~\ref{subsec:briefingeng}: (1) Goal \& Why, (2) What \& Success Criteria, (3) All Needed Context, (4) Implementation Blueprint, and (5) Validation Loop.

\begin{lstlisting}[caption={A BriefingScript for implementing API rate limiting.},label={lst:briefingscript},basicstyle=\ttfamily\scriptsize,frame=single,breaklines=true,escapeinside={(*@}{@*)}]
# BriefingScript: Implement Rate Limiting for REST API
briefing_id: BS-2025-API-012
version: 1.2

# Section 1: Goal & Why
goal:
  objective: "Add rate limiting to prevent API abuse."
  business_value: "Protects infrastructure from DDoS,
                   reduces costs, improves reliability."

# Section 2: What & Success Criteria
success_criteria:
  definition_of_done:
    - "Rate limiter enforces 100 req/min per API key."
    - "Exceeded requests return HTTP 429 with Retry-After."
  invariants:
    - "Rate limiting must be idempotent across instances."
  preconditions:
    - "Valid API key exists in request header."

# Section 3: All Needed Context
context:
  relevant_files:
    - "src/middleware/auth.ts"
    - "src/config/rate-limits.yaml"
  documentation:
    - "docs/architecture/api-gateway.md"
  known_gotchas:
    - "Redis cluster has ~50ms replication lag."
    - "Legacy /v1/* endpoints must remain unlimited."

# Section 4: Implementation Blueprint
blueprint:
  recommended_approach: "Use sliding window with Redis."
  constraints:
    - "Do NOT modify auth middleware directly."
    - "Use existing Redis pool from src/db/redis.ts."
  patterns_to_use:
    - "Strategy pattern for swappable algorithms."

# Section 5: Validation Loop
validation:
  levels:
    - type: "unit"
      description: "Test sliding window with mock Redis."
    - type: "integration"
      description: "Verify 429 responses under load."
    - type: "acceptance"
      description: "Confirm Retry-After per RFC 7231."
\end{lstlisting}

Listing~\ref{lst:briefingscript} provides more than a conventional task ticket. For example, the \texttt{known\_gotchas} field captures project knowledge that teams rarely document (e.g., the replication lag of the Redis cluster), giving the agent the necessary context without requiring it to read the entire codebase. The blueprint specifies the strategy, including which connection pool to reuse and which middleware to leave untouched, while leaving implementation details to the agent. The validation loop specifies how to check the definition of done at the unit, integration, and acceptance levels, making the task's intent testable. The brief also contains a tension between prescribing Redis and warning about its replication lag. By making both constraints explicit, the brief helps the agent identify the conflict and escalate it to a human.

\subsection{Consultation Request Pack (CRP) Example}\label{sec:appendix_crp}

Listing~\ref{lst:crp} illustrates a Consultation Request Pack generated by an agent when it encounters an ambiguity that requires human judgment. In this scenario, the agent working on the BriefingScript from Listing~\ref{lst:briefingscript} identifies a conflict between the specified Redis backend and the known gotcha about replication lag for high-stakes endpoints. The CRP structures the decision point with options, trade-offs, and a recommendation, so a human can intervene efficiently. Option B reduces the risk associated with replication lag by having Redis wait for replica acknowledgments before replying, but it still cannot guarantee strong consistency or prevent all write loss during failover.\footnote{\url{https://redis.io/docs/latest/operate/oss_and_stack/management/replication/}} Hence, the human must decide whether this residual risk is acceptable or whether the task needs a different backend.

\begin{lstlisting}[caption={A Consultation Request Pack (CRP) for an architectural decision.},label={lst:crp},basicstyle=\ttfamily\scriptsize,frame=single,breaklines=true,escapeinside={(*@}{@*)}]
# Consultation Request Pack: Architectural Decision
crp_id: CRP-2025-API-012-001
source_agent: "developer-agent-03"
triggered_by: "BriefingScript BS-2025-API-012"
timestamp: "2025-06-15T14:32:00Z"

# The Decision Point
issue:
  summary: "Conflicting requirements for rate limit backend."
  details: |
    The BriefingScript specifies Redis, but known_gotchas
    warn about replication lag. For /v2/payments (high-stakes),
    even 50ms inconsistency could allow burst abuse.

# Structured Options for Human Decider
options:
  - id: A
    label: "Redis with local cache fallback"
    pros: ["Matches blueprint", "Simple implementation"]
    cons: ["Inconsistency window during failover"]
    estimated_effort: "2 hours"

  - id: B
    label: "Redis with replica acknowledgments (WAIT)"
    pros: ["Replicas confirm each counter update"]
    cons: ["Added latency", "Write loss still possible on failover"]
    estimated_effort: "4 hours"

# Agent's Recommendation
recommendation:
  preferred_option: "B"
  reasoning: |
    Prefer replica acknowledgments if the human accepts
    residual failover risk. Otherwise, revise the backend.
    Set timeout behavior and a latency budget before execution.

# Request for Human
requested_decision: "Confirm Option A or B, or provide
                     alternative guidance."
escalation_target: "Tech Lead or Architect role"
\end{lstlisting}

The CRP implements the escalation mechanism of Agentic Guidance Engineering (Section~\ref{subsec:age}). The agent narrows the problem to a single decision, presents two concrete options with their trade-offs and estimated effort, and recommends one of them with supporting reasoning. Instead of posing an open-ended question and waiting, the agent provides a structured proposal for the human to approve or redirect. The \texttt{triggered\_by} field links the request to the originating BriefingScript, while the \texttt{escalation\_target} field routes it to a person with the appropriate expertise. The MRP in Appendix~\ref{sec:appendix_mrp} references the resolution of this request through a versioned identifier (VCR-2025-API-012-001 v1). In practice, the workflow would store the full decision and its supporting evidence under that identifier.

\subsection{Merge-Readiness Pack (MRP) Example}\label{sec:appendix_mrp}

Listing~\ref{lst:mrp} shows a Merge-Readiness Pack submitted by the agent upon completing the rate limiting implementation from Listing~\ref{lst:briefingscript}. In this scenario, the human approved Option B of the CRP (Listing~\ref{lst:crp}) and accepted its residual failover risk, so the agent enabled replica acknowledgments for the \texttt{/v2/payments} rate limiter. The MRP bundles evidence demonstrating functional completeness, sound verification, SE hygiene, clear rationale, and full auditability, which are the five criteria for merge-readiness defined in Section~\ref{subsubsec:merge_pack}.

\begin{lstlisting}[caption={A Merge-Readiness Pack (MRP) for the completed rate limiting feature.},label={lst:mrp},basicstyle=\ttfamily\scriptsize,frame=single,breaklines=true,escapeinside={(*@}{@*)}]
# Merge-Readiness Pack: Rate Limiting Implementation
mrp_id: MRP-2025-API-012
source_briefing: "BS-2025-API-012 (v1.2)"
agent: "developer-agent-03"
submitted: "2025-06-15T18:45:00Z"

# 1. Functional Completeness
functional_completeness:
  criteria_met:
    - criterion: "100 req/min per API key"
      status: PASS
      evidence: "integration_test_rate_limit.log"
    - criterion: "HTTP 429 with Retry-After header"
      status: PASS
      evidence: "test_429_response.log"
    - criterion: "Invariant: idempotent across instances"
      status: PASS
      evidence: "multi_instance_consistency.log"
    - criterion: "Failure policy per VCR-2025-API-012-001 v1"
      status: PASS
      evidence: "replica_failover_test.log"
  notes: "Option B implemented per VCR-2025-API-012-001 v1:
          replica acknowledgments enabled."

# 2. Sound Verification
verification:
  test_plan: "Per-key window and middleware ordering tests;
              load and 429 checks; replica timeout and failover
              checks required for the accepted failure policy"
  unit_tests: {total: 24, passed: 24, coverage: "92%"}
  integration_tests: {total: 8, passed: 8}
  load_test:
    tool: "k6"
    result: "429 returned at 101st request within 60s"
  acceptance_test:
    description: "Retry-After header format per RFC 7231"
    status: PASS

# 3. SE Hygiene
hygiene:
  linting: {warnings: 0, errors: 0}
  complexity:
    cyclomatic_max: 6
    threshold: 10
    status: PASS
  duplication: "No new duplicated blocks detected"

# 4. Rationale & Communication
rationale:
  approach: |
    Implemented sliding window algorithm with Redis backend.
    Used Strategy pattern for algorithm flexibility.
    Added replica acknowledgments for /v2/payments per
    VCR-2025-API-012-001 v1 (accepts residual failover risk).
  trade_offs: |
    Replica acknowledgments add latency and do not guarantee
    strong consistency. The human accepts the residual risk.

# 5. Auditability
auditability:
  briefing_hash: "sha256:a1b2c3..."
  mentorscript_version: "v2.1.0"
  crp_resolutions: ["VCR-2025-API-012-001 v1 (CRP-2025-API-012-001)"]
  trajectory_id: "trace-2025-06-15-agent03-api012"
  artifacts:
    - "src/middleware/rate-limiter.ts"
    - "src/middleware/rate-limiter.test.ts"
    - "docs/rate-limiting.md"
\end{lstlisting}

A reviewer receiving this pack must decide whether the change is ready to merge. The five blocks of the MRP address this question in the order of the merge-readiness criteria defined in Section~\ref{subsubsec:merge_pack}. The reviewer first assesses whether the change satisfies the brief. The functional-completeness block addresses this by linking each definition-of-done criterion to an evidence file. The block also covers the brief's invariant and the failure policy that the human accepted in the CRP resolution. Next, the reviewer examines how the agent verified its work. The verification block pairs the agent's test plan with the results of unit, integration, load, and acceptance tests, so the reviewer can judge the testing strategy and not only whether the tests passed. The hygiene block demonstrates maintainability through linting, complexity, and duplication checks, while the rationale block explains the approach and its trade-offs, which the reviewer would otherwise need to reconstruct from the raw agent trajectory. Finally, the auditability block connects the delivered code to its history. It references the BriefingScript hash, the MentorScript version, the trajectory, and the versioned CRP resolution, allowing the reviewer to trace the human decision in Appendix~\ref{sec:appendix_crp} to the resulting code.

\end{document}